\documentclass[usegraphicx]{mn2e}
 \usepackage{subcaption}
\usepackage{epsfig}
\usepackage{amsmath}
\usepackage{amssymb}
\usepackage{graphicx}
\usepackage{times,longtable,color}
\usepackage{mathrsfs}

\usepackage{wrapfig}

\newcommand {\apgt} {\ {\raise-.5ex\hbox{$\buildrel>\over\sim$}}\ }
\newcommand {\aplt} {\ {\raise-.5ex\hbox{$\buildrel<\over\sim$}}\ } 
\newcommand {\degree}{$^{\circ}$}

\newcommand{\appropto}{\mathrel{\vcenter{
  \offinterlineskip\halign{\hfil$##$\cr
    \propto\cr\noalign{\kern2pt}\sim\cr\noalign{\kern-2pt}}}}}

\title[The Lense-Thirring timing-accretion plane for ULXs]
{The Lense-Thirring timing-accretion plane for ULXs}

\author[M. Middleton et al.]
{M. J. Middleton$^{1}$, P. C. Fragile$^{2, 3}$, A. Ingram$^{4}$ \& T. P. Roberts$^{5}$\\
\\
1. Department of Physics and Astronomy, University of Southampton, Highfield, Southampton SO17 1BJ, UK\\
2. Department of Physics and Astronomy, College of Charleston, Charleston, SC 29424, USA\\
3. Kavli Institute for Theoretical Physics, Santa Barbara, CA, USA\\
4. Department of Physics, Oxford University, Denys Wilkinson Building, Keble Road, Oxford OX1 3RH, UK\\ 
5. Centre for Extragalactic Astronomy, Durham University, Dept of Physics, South Road, Durham DH1 3LE, UK\\ 
}

\pagerange{\pageref{firstpage}--\pageref{lastpage}} \pubyear{2017}
\long\def\symbolfootnote[#1]#2{\begingroup\def\thefootnote{\fnsymbol{footnote}}\footnote[#1]{#2}\endgroup} 
\def\ga{\mathrel{\hbox{\rlap{\hbox{\lower4pt\hbox{$\sim$}}}{\raise2pt\hbox{$>$}}
}}}

\begin{document}

\topmargin = -0.5cm

\maketitle

\label{firstpage}

\begin{abstract}

Identifying the compact object in ultraluminous X-ray sources (ULXs) has to-date required detection of pulsations or a cyclotron resonance scattering feature (CRSF), indicating a magnetised neutron star. However, pulsations are observed to be transient and it is plausible that accretion onto the neutron star may have suppressed the surface magnetic field such that pulsations and CRSFs will be entirely absent. We may therefore lack direct means to identify neutron star systems whilst we presently lack an effective means by which to identify black hole ULXs. Here we present a possible method for separating the ULX population by assuming the X-ray, mHz quasi-periodic oscillations (QPOs) and day timescale periods/QPOs are associated with Lense-Thirring precession of the inflow and outflowing wind respectively. The precession timescales combined with the temperature of the soft X-ray component produce planes where the accretor mass enters as a free parameter. Depending on the properties of the wind, use of these planes may be robust to a range in the angular momentum (spin) and, for high accretion rates, essentially independent of the neutron star's surface dipole field strength. Our model also predicts the mHz QPO frequency and magnitude of the phase-lag imprinted due to propagation through the optically thick wind; in the case of NGC 5408 X-1 we subsequently infer a black hole mass and high spin. Finally, we note that observing secular QPO evolution over sufficient baselines may indicate a neutron star, as the precession responds to spin-up which is not readily observable for black hole primaries.

\end{abstract}






\section{introduction}

The discovery of X-ray pulsations with periods on the order of seconds in five ULXs (three of which are amongst the brightest of the population, Bachetti et al. 2014; Fuerst et al. 2016; Israel et al. 2017a, 2017b, Carpano et al. 2018; Sathyaprakash et al. 2019) and the detection of a cyclotron resonance scattering feature (CRSF) in M51 ULX-8 (Brightman et al. 2018; Middleton et al. 2019, and potentially in another ULX: Walton et al. 2018) - has led to the inevitable conclusion that some (perhaps most) ULXs harbour neutron stars (NSs), consistent with super-critical disc models with geometrical beaming (King et al. 2001; King \& Lasota 2016; King, Lasota \& Kluzniak 2017) and/or a magnetic pressure supported accretion column (Basko \& Sunyaev 1976; Mushtukov et al. 2015). As the thermal timescale mass transfer rate scales with the companion mass and inversely with the Kelvin Helmholz timescale (whilst its duration in this phase is dictated by the mass ratio), there is no obvious reason why the population of ULXs should not also contain stellar mass black holes (BHs). Indeed indirect arguments based around the nature of the accretion flow and geometrical beaming would support their presence in at least some systems (Middleton \& King 2017) and binary population synthesis studies imply that BH ULXs may even dominate the observed population (Wicktorowicz et al. 2018). However, determining the true relative proportion of neutron stars to black holes is extremely difficult as direct evidence of pulsations or CRSFs has been scant (e.g. Doroshenko et al. 2015) and dynamical mass measurements are remarkably difficult (see Roberts et al. 2011). 

It has been suggested by Middleton et al. (2018) that the long, $\sim$10s of days periods, detected in the ultra-luminous pulsars (ULPs also referred to as PULXs or ULXPs) and seen in other ULXs, could be a consequence of the general relativistic effect of frame-dragging and the resultant Lense-Thirring torque when accreting plasma is misaligned with the equatorial axis of the spinning compact object. The assumption of misalignment is reasonable given that $\sim M$ (the mass of the accreting compact object) has to be transferred for alignment of the neutron star with the system angular momentum (e.g. Nixon \& King 2016) which, even where accretion rates are substantially in excess of the Eddington limit, requires extremely long timescales. Assuming the inflow is super-critical in nature (see Shakura \& Sunyaev 1973; Poutanen et al. 2007), the scale-height of the disc is sufficiently large that the Lense-Thirring torque is communicated by bending waves travelling at $\sim$ half the gas pressure sound speed (see Nelson \& Papaloizou 1999; Fragile et al. 2007) leading to solid-body precession. There is growing observational evidence in support of such precession as it can readily explain both the low frequency QPOs in X-ray binaries at sub-Eddington rates (Ingram et al. 2016, 2017) and the existence of a precessing jet in an X-ray binary accreting at super-Eddington rates (Miller-Jones et al. 2019). In ULXs, in addition to precession of the large scale-height inflow, the radiatively driven, mass loaded wind - disconnected from the inflow - should also precess but on a longer timescale. As the wind is optically thick, this leads to a moving, scattering/absorbing envelope which determines our view of the system. Specifically, the X-ray spectrum is predicted to appear harder and absorption lines - imprinted by outflowing material - weaker when the wind cone points towards the observer (see Poutanen et al. 2007; King 2009; Middleton et al. 2015a, b; Luangtip et al. 2016; Narayan et al. 2017; Dauser et al. 2017; Weng \& Feng 2018), whilst at lower energies, the precession of the wind leads to a predicted anti-correlation between the X-rays and UV (Middleton et al. 2015a; Sonbas et al. 2019). 

Whilst Lense-Thirring precession has been invoked to explain the super-orbital periods in ULPs (Middleton et al. 2018), this is not the only explanation and other possibilities for precession include a slaved-disc (as invoked for SS433: van den Heuvel et al. 1980), radiative warping (Pringle 1996) and precession of the neutron star dipole field (Lipunov \& Shakura 1980). In addition to these, there may be other effects which amplify or dilute the Lense-Thirring torque, notably magnetic torques (e.g. Lai 1999; 2003) and tidal torques from the orbit of the secondary star, both of which were considered by Middleton et al. (2018). The former is predicted to be highly diluted in the case of a thick disc where simulations indicate the viscosity, parameterised by $\alpha$ (Shakura \& Sunyaev 1973) is small (see Jiang et al. 2014, although simulations focussing on the interaction between the disc and external magnetic field are still required to confirm this). The latter is dependent on the mass ratio, orbital period and accretion rate into the system. As can be inferred from equation 8 of Middleton et al. (2018), this is likely to be most relevant for slowly spinning compact objects with short orbital periods or where accretion rate and/or the fraction of accretion luminosity imparted to the wind is very high. We revisit the impact of this torque in light of our developing model in \S 2.1.

A clear prediction of the Lense-Thirring model is that, in addition to a longer precession period of the wind, we may also be able to detect the signature of the precessing inflow buried beneath the wind. We will also show that the periods we might expect from precession of the inflow have frequencies $\sim$mHz and can potentially explain the QPOs observed in some ULXs. As we will show, these precession periods (either alone or as a ratio), when plotted against the temperature of the soft X-rays form planes which may allow the components of the ULX population to be separated.

\section{The model}

The following model is expected to be valid (although see \S 5 for a discussion of the caveats) where ULXs are powered by super-critical accretion rates and, in the case where the primary is a neutron star, the dipole field is not strong enough to prevent the disc locally reaching the Eddington luminosity. This likely requires sub-magnetar dipole field strengths (e.g. Israel et al. 2017; Tsygankov et al. 2017; King \& Lasota 2019) and may naturally result from high rates of accretion onto the neutron star leading to ohmic diffusion and suppression of the surface field (e.g. Bhattacharya 2002, although see also Igoshev \& Popov 2018 for arguments pertaining to long-lived, strong dipole fields). In such a situation, the spherisation radius ($r_{\rm sph}$) - the point where the disc reaches the Eddington limit - is greater than the magnetospheric radius ($r_{\rm m}$). Typically, we can approximate $r_{\rm sph} \approx \dot{m}_{0}r_{\rm isco}$ but a more accurate position, including the role of inwards, radial advection is provided by Poutanen et al. (2007):

\begin{equation}
\frac{r_{\rm sph}}{r_{\rm isco}} = \dot{m}_{0}\left[1.34 - 0.4\epsilon_{\rm wind} + 0.1\epsilon_{\rm wind}^{2} - (1.1-0.7\epsilon_{\rm wind})\dot{m}_{0}^{-2/3}\right], 
\end{equation}

\noindent where $\epsilon_{\rm wind}$ is the fraction of dissipated energy used to launch the wind, $r_{\rm isco}$ is the dimensionless radius of the innermost stable circular orbit (ISCO), set by the spin, and $\dot{m}_{\rm 0} = \dot{M}/\dot{M}_{\rm Edd}$ where $\dot{M}_{\rm Edd} = L_{\rm Edd}/\eta c^{2}$, $L_{\rm Edd}$ is the Eddington luminosity, assumed to be $\approx$1.3$\times$10$^{38}M$/M$_{\odot}$ erg s$^{-1}$ and $\eta$ is the radiative efficiency at the ISCO (which we will assume to be $\approx 1/2r_{isco}$ hereafter). 

The Lense-Thirring precession period of the inflow within $r_{\rm sph}$ (assuming an approximate surface density profile accounting for mass loss: see Middleton et al. 2018) is given by:

\begin{equation}
P_{\rm inflow} = \frac{GM\pi}{3c^{3}a_{*}} r_{\rm sph}^{3} \left[ \frac{1-\left(\frac{r_{\rm in}}{r_{\rm sph}}\right)^{3}}{{\rm ln}\left(\frac{r_{\rm sph}}{r_{\rm in}}\right)}\right],
\end{equation}

\noindent where all radii are expressed in units of the gravitational radius ($r = R/R_{\rm g}$  and $R_{\rm g} = GM/c^{2}$), $a_{*}$ is the dimensionless spin value (= $Jc/GM^{2}$ where $J$ is the angular momentum of the compact object) and $r_{\rm in}$ is the inner edge of the disc. In black hole ULXs the latter will be close to $r_{\rm isco}$, whilst in neutron star ULXs this will be $r_{\rm m}$ (as long as this is larger than the ISCO radius). 

\begin{figure*}
\begin{center}
\includegraphics[trim=0 0 0 0, clip, width=12cm]{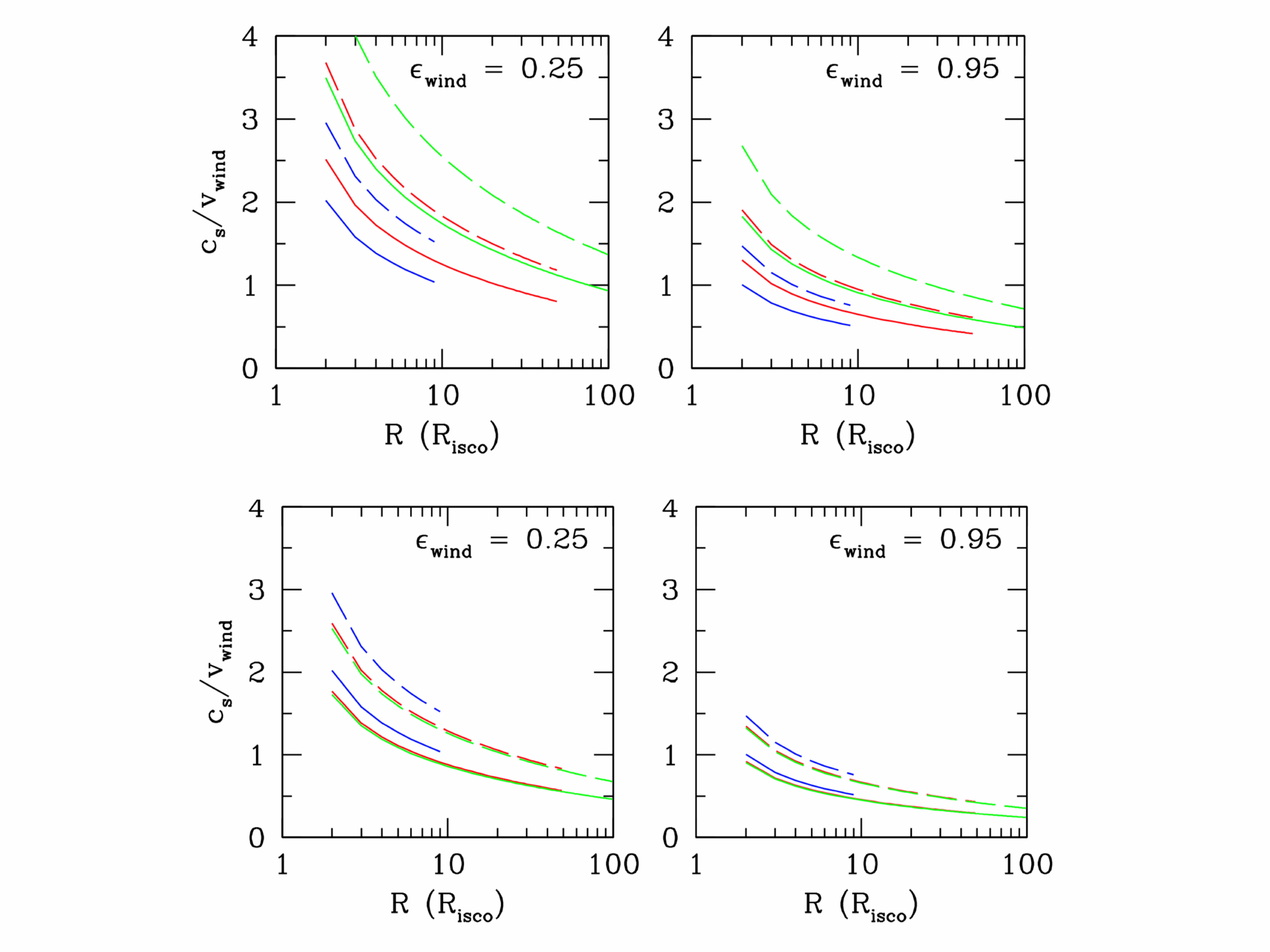}
\end{center}
\vspace{-0.2cm}
\caption{The ratio of isothermal sound speed to wind speed (assumed here to be equal to the local escape speed) for a range of dimensionless spin values (solid line: $a_{*}$ = 0.1, dashed line: $a_{*}$ = 0.998), $\dot{m}_{\rm 0}$ (10 = blue, 50 = red and 100 = green), for $\epsilon_{\rm wind}$ = 0.25 and 0.95, and for $\zeta$ coupled to $\dot{m}_{\rm 0}$ (top panels) and fixed at 2 respectively (bottom panels).} 
\label{fig:l}
\end{figure*}

Middleton et al. (2018) considered the wind to precess on a slower timescale than the inflow from which it is launched assuming conservation of angular momentum. However, Lense-Thirring precession is a pattern of changing angular momentum induced as a consequence of an external torque, and as such this original approach is flawed. Instead a more physical picture would have the wind coupling to the Lense-Thirring torque in the same manner as the inflow at $r_{in}$ but precessing at a slower frequency as a consequence of extending to a larger radius (assuming that it can do so without aligning with the compact object - see Motta et al. 2018). As with the inflow, it is important to determine the radial surface density profile of the wind (i.e. $\Sigma \propto R^{\gamma}$) as this determines the rate of precession. As in Middleton et al. (2018), we can determine the approximate surface density profile from the mass loss rate in the wind: $\dot{M}_{\rm wind}\sim4\pi R^{2}\rho(R) v_{\rm wind}$ where $v_{\rm wind}$ is the wind velocity. Assuming $v_{\rm wind} \propto R^{-1/2}$ (i.e proportional to the local escape velocity) we find that $\dot{M}_{\rm wind} \propto R^{2}\left(\Sigma/H\right)R^{-1/2}$. For the mid-plane height of the disc, $H \propto R$ and given that $\dot{M}_{\rm wind} \appropto R$ (for large $\dot{m}_{0}$ or large $r$, see equation (4) below), we therefore find  $\Sigma(R) \appropto R^{1/2}$ which is the same profile as determined for the inflow (Middleton et al. 2018). We can therefore write a formula analogous to equation (2) but where we assume precession continues out to some larger radius:

\begin{equation}
P_{\rm wind} = \frac{GM\pi}{3c^{3}a_{*}} r_{\rm out}^{3} \left[ \frac{1-\left(\frac{r_{\rm in}}{r_{\rm out}}\right)^{3}}{{\rm ln}\left(\frac{r_{\rm out}}{r_{\rm in}}\right)}\right].
\end{equation}

\noindent In the above, $r_{\rm out}$ is the radius at which the wind becomes optically thin (and, as we show in what follows, we can assume precession terminates here). Following Poutanen et al. (2007), the position of $r_{\rm out}$ can be found from the point at which the perpendicular optical depth ($\tau_{\rm \perp}$) through the wind reaches unity. In turn, the latter can be determined from the mass loss rate measured at a given cylindrical radius $r$, found by integrating the mass flux down to the inner radius (Poutanen et al. 2007; Vasilopolous et al. 2019): 

\begin{equation}
\dot{M}_{\rm wind}(r) = \dot{M}_{\rm Edd}\left(\dot{m}_{\rm 0} - \dot{m}_{\rm in}\right)\frac{(r-r_{\rm in})}{r_{\rm sph}},
\end{equation}

 \noindent where  $\dot{m}_{\rm in}$ is the rate of material making it past the ISCO (i.e onto the neutron star or black hole) in Eddington units given by:

\begin{equation}
\frac{\dot{m}_{\rm in}}{\dot{m}_{\rm 0}} \approx \frac{1-A}{1-A\left(\frac{2}{5}\dot{m}_{\rm 0}\right)^{-1/2}},
\end{equation}

\noindent where $A = \epsilon_{\rm wind}(0.83-0.25\epsilon_{\rm wind}$) (Poutanen et al. 2007). In the case of black holes, we assume the inner disc radius ($r_{\rm in}$) in the above is $r_{\rm isco}$ but in the case of magnetised neutron stars (e.g. Koliopanos et al. 2017; King \& Lasota 2019), this will instead be $r_{\rm m}$, the position of which can be determined from the dipole field strength and accretion rate. We use the formula of Davidson \& Ostriker (1973):

\begin{equation}
R_{\rm m} = 2.9\times10^{8}\dot{M}_{\rm 17}^{-2/7}m_{\rm NS}^{-1/7}\mu_{\rm 30}^{4/7} ~~~[{\rm cm}],
\end{equation}

\noindent where $\dot{M}_{\rm 17}$ is the mass accretion rate in units of $10^{17}$ g s$^{-1}$ at the magnetospheric radius, $m_{\rm NS}$ is the neutron star mass in units of $M_{\odot}$ and $\mu_{\rm 30}$ is the neutron star dipole moment ($BR_{\rm NS}^{3} /10^{30}$ G cm$^{3}$, where $R_{\rm NS}$ is the neutron star radius, assumed to be 10$^{6}$ cm).  Middleton et al. (2018) estimate $\dot{M}$ at $R_{\rm m}$ for a known neutron star ULX by fixing the mass accretion rate at the ISCO to be Eddington limited (even though this radius is not actually reached by the inflow due to truncation at $R_{\rm m}$). We note that this did not include the role of advection and so here we solve numerically for $R_{\rm m}$ by substituting from Poutanen et al. (2007):

\begin{equation}
\dot{M}(R) = \dot{M}_{\rm Edd}\left[\dot{m}_{\rm in} + \left(\dot{m}_{\rm 0} - \dot{m}_{\rm in}\right)\frac{R}{R_{\rm sph}}\right].
\end{equation} 

In an analogous manner to that presented in Vasilopoulos et al. (2019), we obtain the optical depths through the outflow: 

\begin{equation}
\tau_{\rm \perp} (r) \approx \frac{\tau_{0}}{\beta}\frac{\left(\dot{m}_{0}-\dot{m}_{\rm in}\right)}{r_{\rm sph}}\frac{\left({r - r_{\rm in}}\right)}{\sqrt{\left(\frac{r}{r_{\rm isco}}\right)}},
\end{equation} 
 
\noindent for $r \le r_{\rm sph}$. And

\begin{equation}
\tau_{\rm \perp} (r) \approx \frac{\tau_{0}}{\beta}\frac{\left(\dot{m}_{0}-\dot{m}_{\rm in}\right)}{\sqrt{r_{\rm sph}/r_{\rm isco}}}\frac{\left(r_{\rm sph} - r_{\rm in}\right)}{r},
\end{equation} 

\noindent for $r > r_{\rm sph}$. From these we also obtain $\tau_{\rm \parallel}$: 
 
\begin{equation}
\tau_{\rm \parallel} (R) = \frac{1}{\zeta} \int_{R}^{R_{\rm out}}{\tau_{\rm \perp} \frac{dR'}{R'}}.
\end{equation} 
 
\begin{figure*}
\begin{center}
\includegraphics[trim=20 40 20 0, clip, width=16cm, angle=0]{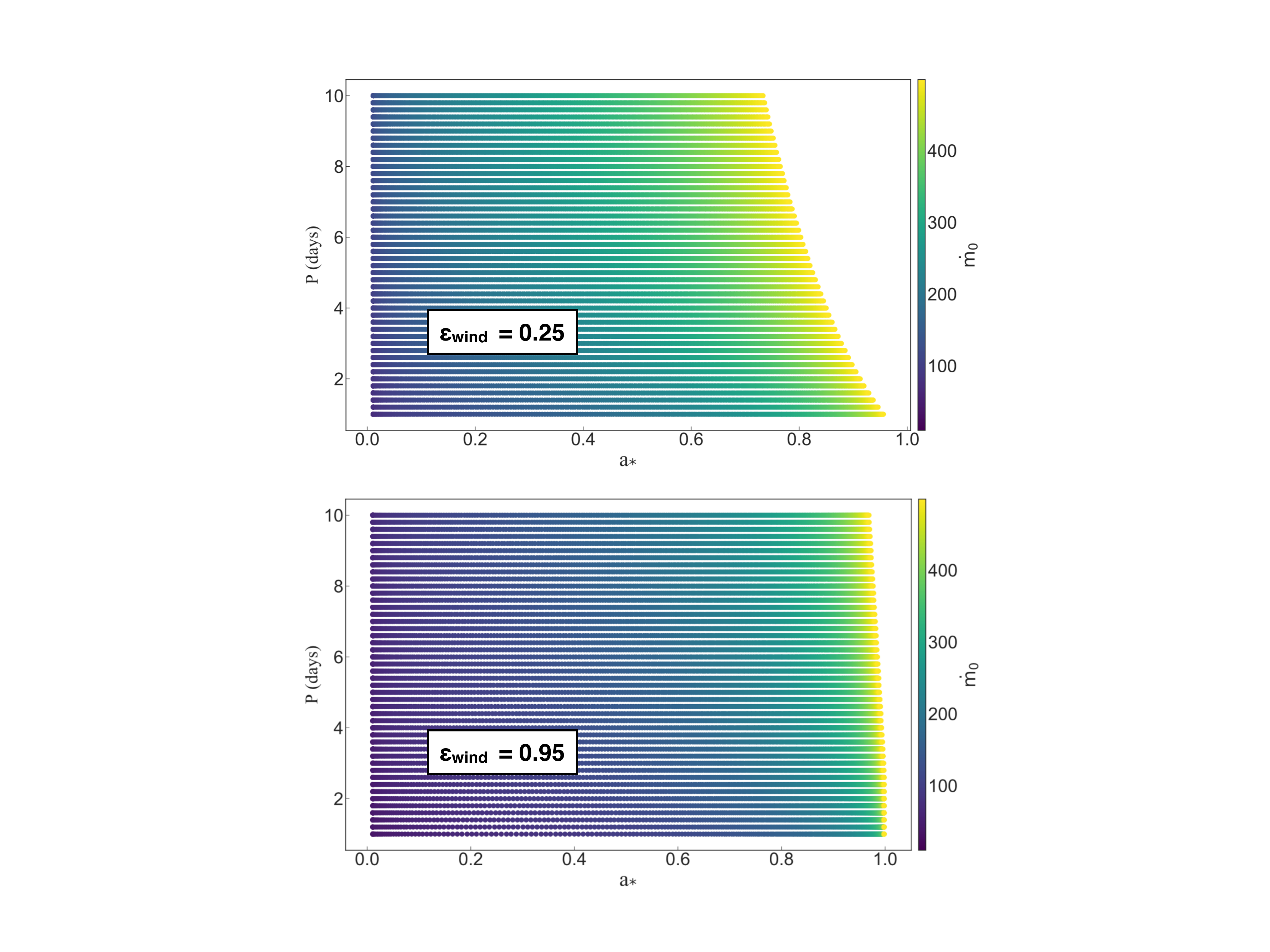}
\end{center}

\caption{Parameter space for black holes (of mass 10 M$_{\odot}$) showing the minimum spin required for the Lense-Thirring torque to dominate over the tidal torque for a given $\dot{m}_{0}$, orbital period $P_{d}$, and with $q$ set at 1.5.} 
\label{fig:l}
\end{figure*}
 
\noindent In the above, $\zeta$ is the cotangent of the opening angle of the wind cone (i.e. H/R$_{\rm wind}$) which, for moderate super-Eddington rates, we expect to be $\approx 2$ based on radiative magneto-hydrodynamic (RMHD) simulations (e.g. Sadowski et al. 2014, although, in reality, $\zeta$ likely increases with $\dot{m}_{0}$: Jiang et al. 2017) and $\beta$ is the ratio of asymptotic wind velocity ($v_{\rm wind}$) relative to the Keplerian velocity at $r_{\rm sph}$. When making predictions about precession (e.g. Middleton et al. 2018), we can source these relevant wind parameters from a mixture of observation (e.g. Pinto et al. 2016; 2017) and simulation (e.g. Sadowski et al. 2014; Jiang et al. 2014, 2017) and note that, in future, the combination of the two should provide considerably improved values. In the above formulae, $\tau _{0}$ is a constant, related to the spin through:
 
\begin{equation}
\tau _{0} = \frac{\dot{M}_{\rm Edd}\kappa\sqrt{r_{\rm isco}}}{4 \pi c R_{\rm isco}},
\end{equation}

\noindent where, assuming that the temperature of the plasma in the wind is not dissimilar from that typically associated with the flow in ULXs (see Middleton et al. 2015a), the opacity, $\kappa$, can be assumed to be dominated by electron scattering (although we return to this in \S 5). As $\dot{M}_{\rm Edd} \propto r_{\rm isco}$ and $R_{\rm isco} \propto r_{\rm isco}$ we then find that $\tau _{0} \propto \sqrt{r_{\rm isco}}$. As $\tau _{0} \approx 5$ for  $r_{\rm isco} = 6$ (Poutanen et al. 2007), we proceed to use:  

\begin{equation}
\tau_{0} \approx 5 \sqrt{\frac{r_{\rm isco}}{6}}.
\end{equation}

By setting $\tau_{\rm \perp}$ = 1 (for $r > r_{\rm sph}$), we then arrive at:

\begin{equation}
r_{\rm out} \approx \frac{\tau_0}{\beta}\frac{\dot{m}_{0} - \dot{m}_{\rm in}} {r_{\rm sph}}\left(r_{\rm sph} - r_{\rm in}\right) r_{\rm isco}\sqrt{r_{\rm sph}}.
\end{equation}

As $r_{\rm sph}$ in units of $r_{\rm isco}$ is proportional to $\dot{m}_{0}$, which in turn is proportional to $1/r_{\rm isco}$, it can be seen that the position of the photospheric radius is essentially independent of the spin as we would expect (but clearly not independent of the dipole field strength when this is large - see also Vasilopoulos et al. 2019). Hereafter we assume the position of the photosphere is given by equation (13), however, as pointed out by Vasilopoulos et al. (2019), it is important to note that for values of $\xi$ larger than $\approx 1$, the position of $r_{\rm out}$ determined from $\tau_{\rm \parallel}$ differs from the above due to the non-spherical nature of the photosphere.  

 

\begin{figure*}
\begin{center}
\includegraphics[trim=20 40 20 0, clip, width=16cm, angle=0]{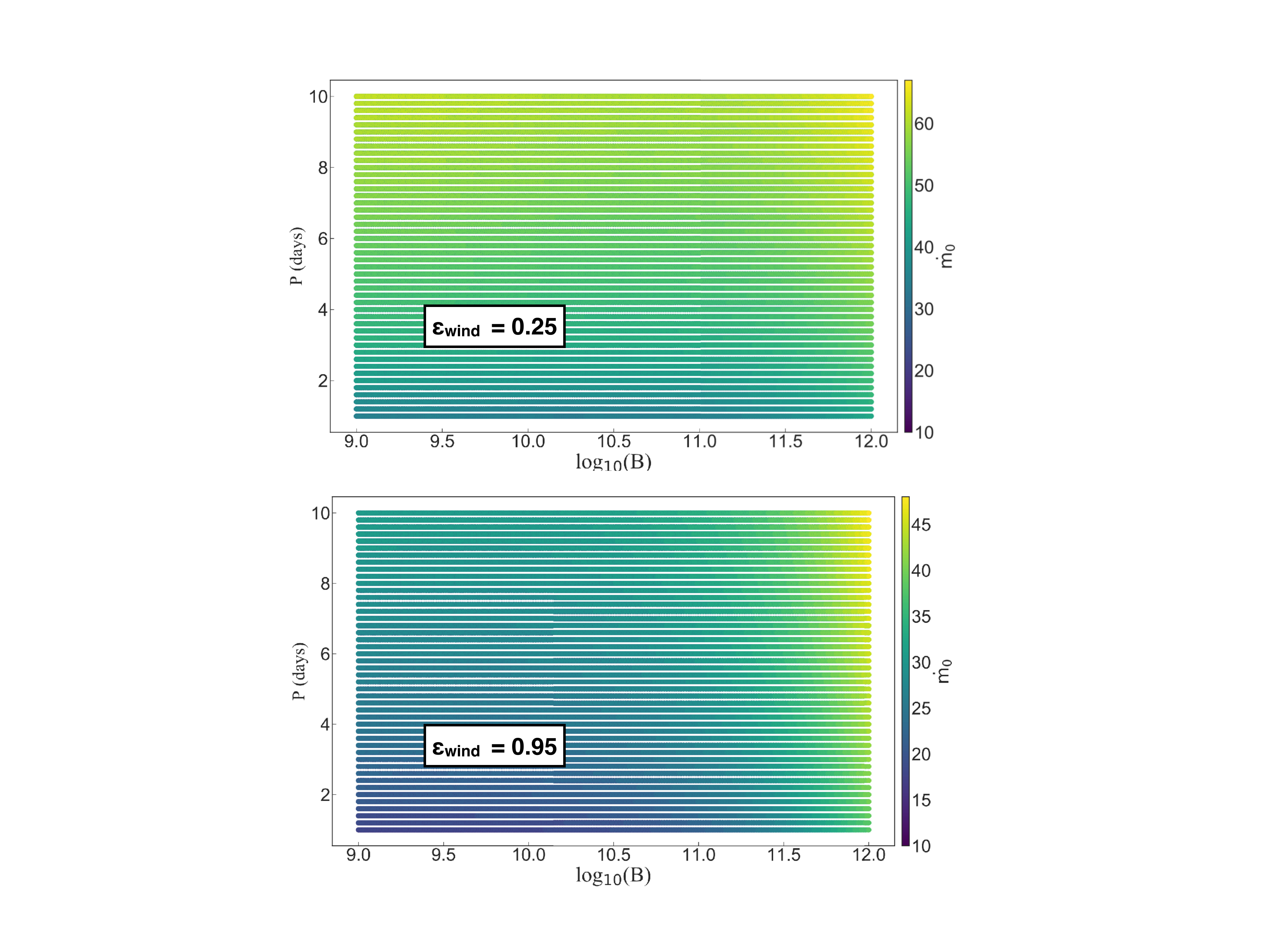}
\end{center}
\caption{Parameter space for neutron stars (of mass 1.4 M$_{\odot}$) showing the maximum $\dot{m}_{0}$ which still allows the Lense-Thirring torque to dominate over the tidal torque for a given dipole field strength, orbital period $P_{d}$, and with $q$ set at 10 and $a_{*}$ set at 0.001.} 
\label{fig:l}
\end{figure*}

Central to our model is the ability of the flow to globally precess, which requires that the Lense-Thirring torque be communicated to all parts of the disk or wind faster than the resulting twisting can propagate through it. This requires that the sound crossing time be less than the local precession timescale. However, if the wind is traveling outward supersonically, then the twisting due to the torque at its base cannot be communicated up through the wind. By assuming the flow to be radiation pressure supported ($P_{\rm rad} = F_{\rm rad}/\sigma = F_{\rm grav}/\sigma = GMm_{\rm p}/R^{2}\sigma$ where $\sigma$ is the Thompson scattering cross-section and $m_{\rm p}$ is the proton mass), we are able to explore whether the wind is sub or super-sonic by estimating the ratio of the isotropic sound speed ($c_{\rm s} = \sqrt{P_{\rm rad}/\rho}$) to the local escape speed (presumably this being close to that of the wind):

\begin{equation}  
\frac{c_{\rm s}}{v_{\rm wind}} = \sqrt{\frac{m_{\rm p}}{2\rho R\sigma}},
\end{equation}

\noindent where from conservation of mass:

\begin{equation}
\rho = \frac{\dot{M}_{\rm wind}}{\zeta4\pi R^{2}v_{\rm wind}} = \frac{\dot{M}_{\rm wind}}{\zeta4\pi \sqrt{2GMR^{3}}}.
\end{equation}

\begin{figure*}
\begin{center}
\includegraphics[trim=20 140 20 140, clip, width=15cm]{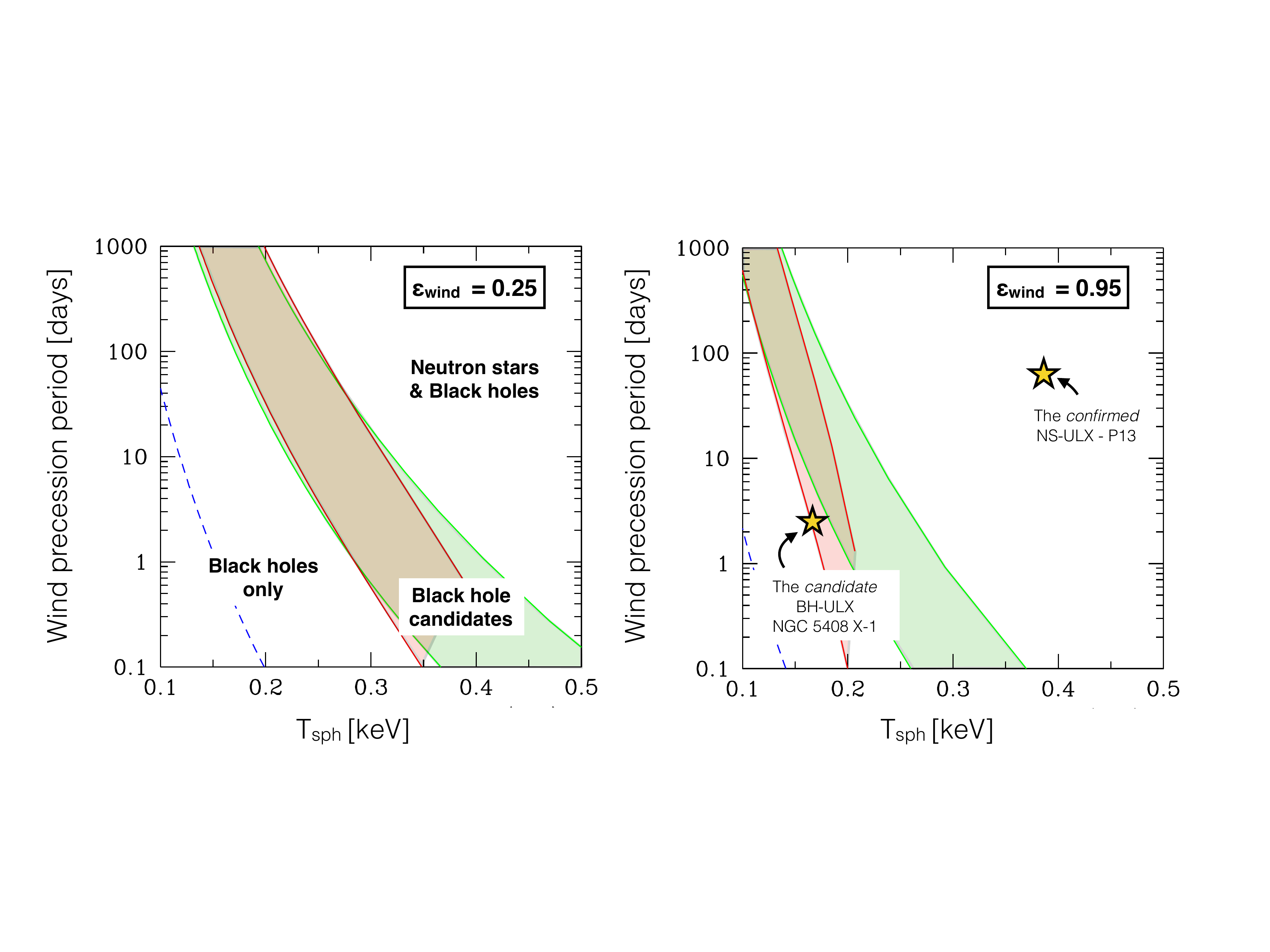}
\end{center}
\vspace{-0.2cm}
\caption{An example of the simplest Lense-Thirring timing-accretion plane for ULXs, where we plot the wind precession period against the observed temperature of the soft X-ray component (for $\epsilon_{\rm wind}$ = 0.25 (left), 0.95 (right), and $f_{\rm col} = 2$). The plot highlights regions where neutron stars and black holes can be found. In green are the limits for neutron stars with dipole field strengths of 10$^{9}$ G with spin values of 0.01 (right-most boundary and a highly conservative upper limit for HMXBs of which ULXs are a sub-class) and 0.3 (left-most boundary and the current observational limit for NSs in binaries: Miller \& Miller 2015). In red we show the analogous case for neutron stars with dipole field strengths of 10$^{12}$ G. Between the respective boundaries we expect to find black hole ULX candidates and to the far-left, only black hole ULXs should be found. For illustrative purposes, we also plot (as a blue dashed line), the parameter bounds for a 50 M$_{\odot}$ black hole with maximal spin and two examples, NGC 7793 P13, a known neutron star ULX (F{\"u}rst et al. 2016; Israel et al. 2017a) and NGC 5408 X-1, which -- as we will show -- is a strong candidate for hosting a black hole.} 
\label{fig:l}
\end{figure*}

Performing the requisite substitutions then allows us to evaluate the ratio of the sound speed to escape speed for a range of input parameters which is mass independent. We assume two cases, one where $\zeta$ = 2 to be consistent with RMHD simulations at moderate super-critical rates (Sadowski et al. 2014) and one where $\zeta$ increases with mass accretion rate, as has been observed to occur in simulations at highly super-critical rates ($\dot{m}_{\rm 0} \sim$100, e.g. Jiang et al. 2017). As we do not yet have a firm scaling relationship, in the case of the latter, we simply assume one based on beaming arguments (King 2009):

\begin{equation}
\zeta = {\rm tan}\left(\frac{\pi}{2} - {\rm acos}\left[1-\frac{73}{\dot{m}_{0}^{2}}\right]\right),
\end{equation} 

\noindent which we set to have a lower limit of $\zeta$ = 2.

As $\dot{M}_{\rm wind}$ is determined from $\dot{m}_{\rm in}$ (equation 4) which depends on $\epsilon_{\rm wind}$, we require input values for this physical parameter. Observational estimates for $\epsilon_{\rm wind}$ imply very high values (even for low covering fractions: Pinto et al. 2016) yet 3D RMHD simulations would imply considerably smaller values (see Jiang et al. 2014). As this issue remains unresolved, we assume two limiting values of $\epsilon_{\rm wind}$: 0.25 and 0.95. We perform calculations for spin values of 0.1 and 0.998 (noting that the highest spin observed in any NS system lies at $a_{*}$ = 0.3: Miller \& Miller 2015), and $\dot{m}_{\rm 0}$ =10, 50 and 100. The ratio of the sound speed to escape speed is plotted in Figure 1 and implies that there is a substantial region of the wind which may be sub-sonic and therefore can presumably communicate the bending wave unhindered. In the case of a neutron star with B $> 10^{9}$ G, the density of the wind is naturally lower, allowing for a more sub-sonic outflow at a given radius.

We note that our calculations have assumed a simple averaged density in eqn (15) whilst, in reality, the structure of the wind will be far more complex (e.g. Takeuchi et al. 2013). We also note that equation (13) is valid for $r \le r_{\rm sph}$; at larger radii the wind is assumed to have the escape speed at $r_{\rm sph}$ (see Poutanen et al. 2007) but the density drops such that, if the wind is subsonic at $r_{\rm sph}$ it is likely to be subsonic to larger radii as well. However, by definition, at $r_{\rm out}$ the wind becomes optically thin to electron scattering (Pountanen et al. 2007) and radiation pressure is no longer effective (leaving only gas and magnetic pressure). As a consequence, at $r_{\rm out}$, the outflow is far more likely to be {\it super}-sonic, regardless of the wind at smaller radii; as a result, global precession will likely cease. This then justifies our use of this limiting radius in equation (3) for the precession timescales of the wind. There is of course uncertainty in this picture which requires GRMHD simulations of the misaligned super-critical flow (which will be performed in the near future), however, we proceed under the assumption that the wind can effectively couple to the Lense-Thirring torque. In addition, in light of present uncertainty in how the opening-angle couples to $\dot{m}_{\rm 0}$, hereafter we assume a fixed value of $\zeta$ = 2.

 \subsection{Impact of binary torques}

As discussed in Middleton et al. (2018), the impact of the tidal (binary) torque may be relevant under certain conditions. The ratio of tidal to Lense-Thirring torque at the location of $r_{\rm out}$ is:

\begin{equation}
\frac{\tau_{\rm tidal}}{\tau_{\rm LT}} = \frac{m_{*}r_{\rm out}^{7/2}}{d^{2}a_{*}m^{5/2}{\rm ln}\left(\frac{r_{\rm out}}{r_{\rm in}}\right)},
\end{equation} 

\noindent where $m_{*}$ and $m$ are the companion and primary mass in M$_{\odot}$ respectively, and $d$ is the orbital separation (in R$_{\rm g}$) given by:

\begin{equation}
d = 2.9\times10^{9}m^{1/3}(1+q)^{1/3}P_{\rm d}^{2/3}\frac{c^{2}}{GM},
\end{equation} 

\noindent where $q = m_{*}/m$ and $P_{\rm d}$ is the orbital period in days. Based on the formulae above, we present illustrative bounds where $\tau_{\rm tidal}/\tau_{\rm LT} \le 1$ in Figures 2 and 3 for the case of a 10 M$_{\odot}$ black hole -- where we show the lower limit on the spin for a given $\dot{m}_{0}$ -- and for a neutron star (of mass 1.4 M$_{\odot}$ with a fixed spin of 0.001) -- where we show the maximum $\dot{m}_{0}$ for a given dipole field strength. In the case of the black hole system, we assume $q$ = 1.5 (which allows for mass transfer on a thermal timescale) and $q$ = 10 for the case of a neutron star (see e.g. Motch et al. 2014; Heida et al. 2019).

\subsection{A Lense-Thirring timing-accretion plane}

\begin{figure*}
\begin{center}
\includegraphics[trim=20 140 20 140, clip, width=15cm]{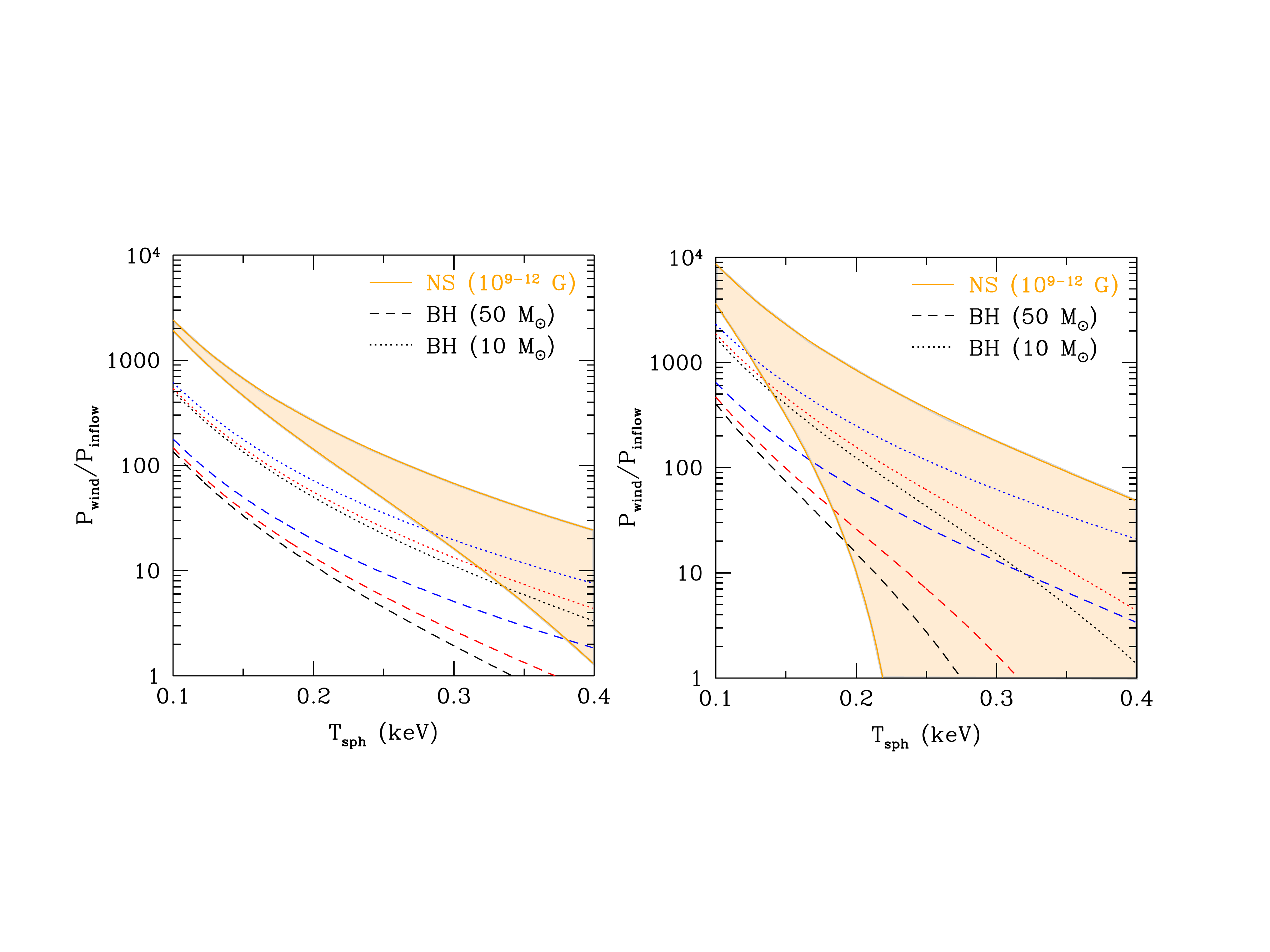}
\end{center}
\vspace{-0.2cm}
    \caption{Plots showing the ratio of $P_{\rm wind}/P_{\rm inflow}$ versus $T_{\rm sph}$ for black holes and neutron stars of various dipole field strengths for $\beta$ = 1.4, $f_{\rm col}$ = 2, and for $\epsilon_{\rm wind}$ = 0.25 (left) and 0.95 (right). We show the corresponding values for black holes with $a_{*}$ = 0.001 (black), 0.5 (red) and 0.998 (blue) and for neutron stars with dipole field strengths of 10$^{9}$ (upper orange line) and 10$^{12}$ G (lower orange line).} 
    \end{figure*}

\subsubsection{1. A simple plane}

Equation (3) (and indeed 2) already provide a means by which the population of ULXs may be somewhat separated. Neutron stars are limited to masses $<$ 2M$_{\odot}$ (Demorest et al. 2010) and observationally to $a_{*} <$ 0.3 (Miller \& Miller 2015; although theoretically, somewhat higher values may be reached: Miller, Lamb \& Cook 1998). Although such high spins may be reached in LMXBs, in HMXBs and ULXs, the spin frequency is $\sim$ 1~Hz (see http://www.iasfbo.inaf.it/\~mauro/pulsar\_list.html) and $a_{*}$ should be $\ll$ 0.01. Irrespective of the spin, the neutron star surface prevents the inner edge of the disc sitting far below 6~$R_{\rm g}$ whilst in black holes (which can have masses up to at least $\approx$50 M$_{\odot}$: Abbott et al. 2015), the ISCO can recede to its minimal value of $\approx$1.25~$R_{\rm g}$ for a maximal spin of $a_{*}$ = 0.998. As equations (2) and (3) are sensitive to the ratio of $M/a_{*}$, it is apparent that the physical differences between the species could potentially lead to differences in their precession timescales. Equations (2) and (3) also indicate that the precession timescale is a function of $\dot{m}_{0}$ (via equations (1) and (13)), which, in the classical super-critical model, is related to the peak of the black-body-like emission in the soft X-rays when associated with emission from $r_{\rm sph}$ (with the observed radiation necessarily propagated through the wind). The formula for this colour temperature is estimated by Poutanen et al. (2007) to be:

\begin{equation}
T_{\rm sph} \approx 1.5 f_{\rm col}\dot{m}_{\rm 0}^{-1/2}m^{-1/4}\left(\frac{6}{r_{\rm isco}}\right)^{1/2} ~~~~ [{\rm keV}],
\end{equation}

\noindent and is dependent on the colour temperature correction factor from scattering and absorption ($f_{col}$), the mass of the compact object, $m$ (in units of solar masses), and we have included the dependence on the spin through $r_{\rm isco}$. As pointed out by Poutanen et al. (2007) who first derived the formula for $T_{\rm sph}$, the temperature should be reduced by a factor of $(1-\epsilon_{\rm wind})^{1/4}$ to account for energy imparted to the wind. 

We consider a canonical 1.4 M$_{\odot}$ neutron star with a weak dipole field of 10$^{9}$ G (such that typically $r_{\rm m} \le r_{\rm isco}$), and a stronger dipole field of 10$^{12}$ G to be consistent with those values reported by some studies of the ULPs and candidate NS ULXs (see Christodoulou et al. 2016; F{\"u}rst et al. 2016; King \& Lasota 2016; King et al. 2017; Koliopanos et al. 2017; Carpano et al. 2018; Vasilopoulos et al. 2018; King \& Lasota 2019; Middleton et al. 2019, although see also Eksi et al. 2015; Dall Osso et al. 2015; Tsygankov et al. 2016). We assume limits of $a_{*} = 0.01$ (HMXBs) and $a_{*} = 0.3$ (the present observational limit for any NS binary: Miller \& Miller 2015), $\epsilon_{\rm wind}$ = 0.25 and 0.95, $\beta$ = 1.4 (such that $v_{\rm wind}$ is the local escape velocity, $\beta\sqrt{GM/R}$), $\zeta$ = 2, $f_{\rm col} = 2$, and plot the parameter space of wind precession period (we use the wind period as this should leave the strongest imprint on the X-ray lightcurve) against $T_{\rm sph}$. We also test for the case where we have black holes with masses up to 50 M$_{\odot}$ in accordance with the largest merger mass so far detected by LIGO (Abbott et al. 2016) - although the end-product black hole from a merger is not expected to have a companion from which to accrete. Figure 4 demonstrates that there are clear regions in which only black holes should be found either as candidates (where realistically neutron star ULXs are unlikely to be found given their expected spin values of $<$ 0.01) or where observations imply neutron stars will not be found (with spins $>$ 0.3). To illustrate the latter, we have highlighted the precession period and temperature for a 50 M$_{\odot}$ black hole at maximal spin (blue dashed line in Figure 4) and included two ULXs where precession periods and temperatures have been reported - NGC 7793 P13 (a known neutron star ULX: F{\"u}rst et al. 2016; Israel et al. 2017a) and NGC 5408 X-1 which, as we will argue later is a candidate for harbouring a black hole. 

In terms of observations and our ability to separate out the ULX population, the Lense-Thirring timing-accretion plane illustrates a potentially powerful yet simple technique, as all that would be required to locate black hole ULXs would be the precession period (easily accessible in the X-rays and possibly other wavebands, e.g. Middleton et al. 2015a; Sonbas et al. 2019) and the temperature of the soft X-ray component which is easy to obtain given the effective area of most X-ray detectors at such energies. We note however, that, in constructing the plane shown in Figure 4, we have proceeded under the assumption that the wind is fully disconnected from the inflow which is unlikely to be a true physical reflection of the system (due to e.g. threading of magnetic fields), whilst the precise locations of the various regions identified in Figure 4 are naturally subject to uncertainties (e.g. in the wind parameter values and the location of $r_{\rm out}$) - both of these potential issues will be probed in the near future via simulations. 

\begin{figure*}
\begin{center}
\includegraphics[trim=200 100 160 100, clip, width=10cm]{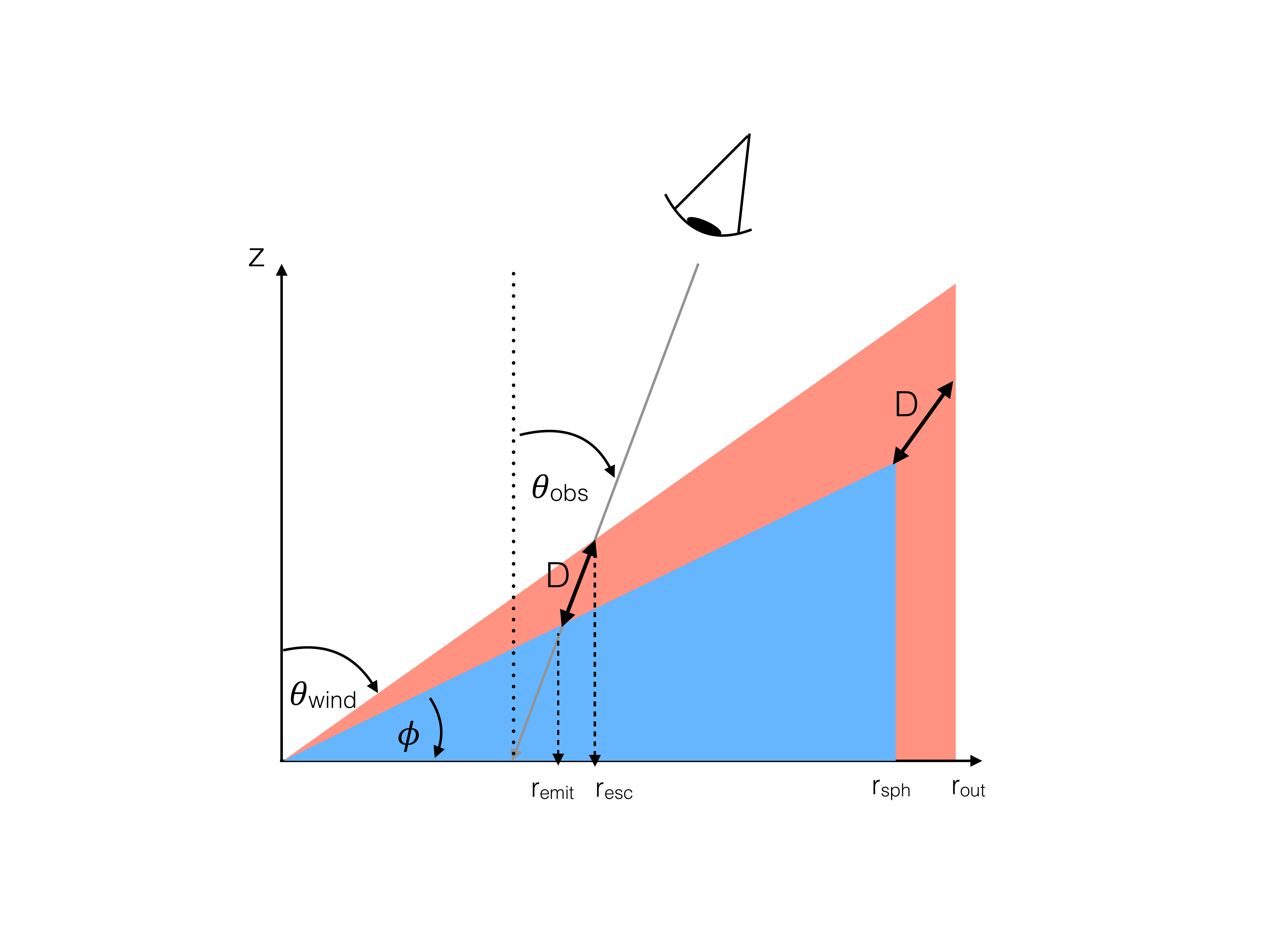}
\end{center}
\vspace{-0.2cm}
\caption{Schematic plot of the model. Photons escape from the blue component of the disc (with a scale-height of $\sim$0.6: Lipunova 1999) which precesses on a timescale set by the accretion rate, compact object mass, spin and surface density profile (see Fragile et al. 2007; Middleton et al. 2018). These photons are emitted at $r_{\rm emit}$ pass through a distance $D$ of wind material and escape at radius $r_{\rm esc}$ (noting that $D$ can start to reduce when $r_{\rm emit}$ becomes large, as indicated at $r_{\rm sph}$). The passage of these photons will lead to a dilution in variability power and a lag at the QPO frequency between well-separated energy bands as the optical depth of the wind and distance through it are both a function of radius (and inclination).} 
\label{fig:l}
\end{figure*}

\subsubsection{The ratio plane}

Whilst the above approach might allow us to identify candidate black holes ULXs, some or even most could in principle lie above the topmost dividing line in Figure 4 (e.g. if born with low natal spin) and be indistinguishable in the X-rays from neutron stars, where pulsations and CRSFs may be absent. Concordantly, the plane shown in Figure 4 provides no means to identify NS ULX candidates; however, the {\it ratio} of the two precession frequencies (wind and disc) may allow the populations to be separated further: 

 \begin{equation}
 \frac{P_{\rm wind}}{P_{\rm inflow}} \approx \left(\frac{r_{\rm out}}{r_{\rm sph}}\right)^{3} \left[\frac{1-\left(r_{\rm in}/r_{\rm out}\right)^{3}}{1-\left(r_{\rm in}/r_{\rm sph}\right)^{3}}\right]  \left[\frac{{\rm ln}\left(r_{\rm sph}/r_{\rm in}\right)}{{\rm ln}\left(r_{\rm out}/r_{\rm in}\right)}\right].
 \end{equation}
 
\noindent The formula indicates that, whilst the precession frequencies themselves depend on physical quantities such as the mass and spin, the ratio of the two frequencies does not depend on the mass. As we will show, there is only a weak dependence on the spin, and the impact of dipole field strength in the case of NS ULXs depends on $\dot{m}_{0}$ and $\epsilon_{\rm wind}$ (assuming that the wind parameters are not heavily dependent on the spin or dipole field strength). 




Whilst a limit on $\dot{m}_{\rm 0}$ can be placed on a given ULP by considering the rate of spin-up (e.g. King, Lasota \& Kluzniak 2017), as can be seen in equation (19), it can also be obtained from $T_{\rm sph}$. If we assume that the properties of outflows do not vary greatly, such that on average $\epsilon_{\rm wind}$ and $\beta$ are the same for black hole and neutron star ULXs for the same Eddington scaled mass accretion rate, then, by plotting $P_{\rm wind}/P_{\rm inflow}$ versus $T_{\rm sph}$, we might expect to see two populations indicating the neutron star and black hole ULXs. 

We again assume two limiting values of $\epsilon_{\rm wind}$: 0.25 and 0.95, $\beta$ = 1.4 and $f_{\rm col}$ = 2. In Figure 5 we plot the resulting theoretical distribution of $P_{\rm wind}/P_{\rm inflow}$ for black holes with masses of 50 M$_{\odot}$ and 10 M$_{\odot}$ (for a range of spins), and neutron stars with a canonical mass of 1.4 M$_{\odot}$. For the neutron stars, we investigate for dipole field strengths of 10$^{9}$ and 10$^{12}$G, and assume for simplicity that the NS spin is low enough such that the ISCO sits at 6~R$_{\rm g}$ (we naturally restrict $R_{\rm m}$ to being at or larger than this radius). As can be inferred from the figure, there are once again clear regions in which neutron stars are not expected to appear, however at large values of $\epsilon_{\rm wind}$, the distinction between regions is restricted to $T_{\rm sph} \lesssim 0.15$ keV. Should an object fall in this region then, from Figure 4, we would already expect the source to harbour a black hole for the vast majority of precession timescales shorter than years. The ability to use this technique effectively will therefore be helped by having an accurate value for $\epsilon_{\rm wind}$ and, where this is large, constraints on the dipole field strength via indirect arguments (e.g. Middleton et al. 2019; Mushtukov et al. 2019). Where these are obtained (or if $\epsilon_{\rm wind}$ is small), we can hope to use this second method to effectively separate out the NS from the BH ULXs.

\subsubsection{Secular evolution}

Whilst our identfied `planes of accretion' shown in Figures 4 and 5 should allow us to start separating the compact object populations in ULXs, we note that, where a neutron star is present with a high magnetic field and/or high accretion rate (such that the accretion torque is high), the spin-up should lead to a potentially observable evolution in the precession periods over sufficiently long observing baselines. This is true whether the precession is driven by the Lense-Thirring effect as in our model or instead driven by precession of the magnetic dipole (Lipunov \& Shakura 1980; Mushtukov et al. 2017) as the dimensionless spin enters into both. As the spin-up rate is much slower when the compact object is instead a black hole (see Fragos \& McClintock 2015), such secular evolution in the precession period(s) will likely indicate the presence of a neutron star in a given source. It is important to note that variations in accretion rate will add a background `noise' against which evolution of the periods will be imprinted and this must be carefully accounted for when searching for secular trends. 

\section{Identifying precession of the inflow}

So far we have speculated that Lense-Thirring precession of the wind drives the long (typically 10s of days) timescale period we observe in ULPs and a number of ULXs (with as-yet unidentified primaries). However, to verify that such precession is indeed occurring, it is clearly important to find the signature of the precessing inflow below the wind. As with precession of the wind, this signal should appear as a QPO (quasi-periodic as any fluctuations in the accretion rate in the outer disc, $\dot{m}_{\rm 0}$, will change the location of $r_{\rm sph}$ and the precession period via equation 2), but any escaping photons from the disc must be scattered by the outflow lying above the disc. In the simplest case, we might expect to see the convolution of the intrinsic QPO with a rectangular impulse response function with width given by the light travel time through the outflow = $\tau^{2}/c\kappa\rho = \tau D/c$ such that:

\begin{equation}
\begin{split}
\prod(t) & = 1   ~~~~~    t \le  \tau D/c \\
 & = 0   ~~~~~   t > \tau D/c,
\end{split}
\end{equation}

\noindent where $\tau$ is the total optical depth along the line-of-sight through the wind (of opacity $\kappa$ and density $\rho$) averaged over a precession cycle, and $D$ is the distance through the outflow to the disc. As we indicate in Figure 6, $D$ extends from the radial position where the photon is emitted ($r_{\rm emit}$) to that where it escapes from the wind ($r_{\rm esc}$) and depends on $\theta_{\rm obs}$, the angle to our line-of-sight (which we take to be averaged over a QPO period, noting that the instantaneous inclination will change with the QPO phase). We assume that $\tau$ is the combination of the optical depths in the horizontal (perpendicular to the instantaneous disc axis) and radial (parallel to the instantaneous disc axis) components of the outflow optical depth ($\tau_{\rm \perp}$ and $\tau_{\rm \parallel}$ respectively):

 \begin{equation}
\tau =  \tau_{\perp}(r_{\rm esc}) + \tau_{\parallel}(r_{\rm esc}) - \tau_{\parallel}(r_{\rm emit}).
 \end{equation} 
 
\noindent We note that, once again, we are making the simplifying assumption that the outflow is laminar whilst this is almost certainly not the case (see Takeuchi et al. 2013) and we ignore any torques induced by the propagation of the radiation; in future we will address these issues directly via use of GRMHD simulations. The variability power of the QPO from the precessing inflow will be diluted as a consequence of the signal's passage through the wind. A similar situation was also considered by Mushtukov et al. (2019), where variability is suppressed by scatterings in an optically thick accretion curtain (Mushtukov et al. 2017). The effect of this propagation is equivalent to a convolution of the QPO signal with the impulse response function in the time domain, and the final emergent (observed) power scaling as the squared Fourier transform of this (the transfer function):

\begin{equation}
\prod(t) \xrightarrow{\mathscr{F}} {\rm sinc}(\pi\nu\tau D/c), 
\end{equation}

\noindent where $\nu$ = $\nu_{QPO}$ at the QPO centroid frequency (and can be obtained from equation (2)). 

\begin{figure*}
\begin{center}
\includegraphics[trim=20 230 40 300, clip, width=12cm]{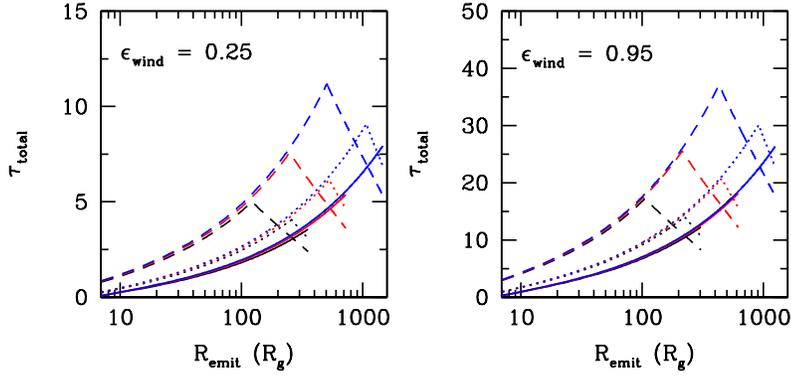}
\end{center}
\caption{Optical depth through the wind versus emitting radius (in units of $R_{\rm g}$), for a black hole system viewed at a given inclination angle ($\theta_{\rm obs}$, averaged over the precessional phase of the inflow) and as a function of $\dot{m}_{\rm 0}$ and $\epsilon_{\rm wind}$. Black lines indicate $\dot{m}_{\rm 0}$ = 50, red lines indicate $\dot{m}_{\rm 0}$ = 100 and blue lines indicate $\dot{m}_{\rm 0}$ = 200. The solid lines indicate $\theta_{\rm obs}$ = 0\degree, dotted lines are $\theta_{\rm obs}$ = 10\degree and dashed lines are $\theta_{\rm obs}$ = 20\degree.} 
\label{fig:l}
\end{figure*}

\begin{figure*}
\begin{center}
\includegraphics[trim=80 80 50 50, clip, width=16cm]{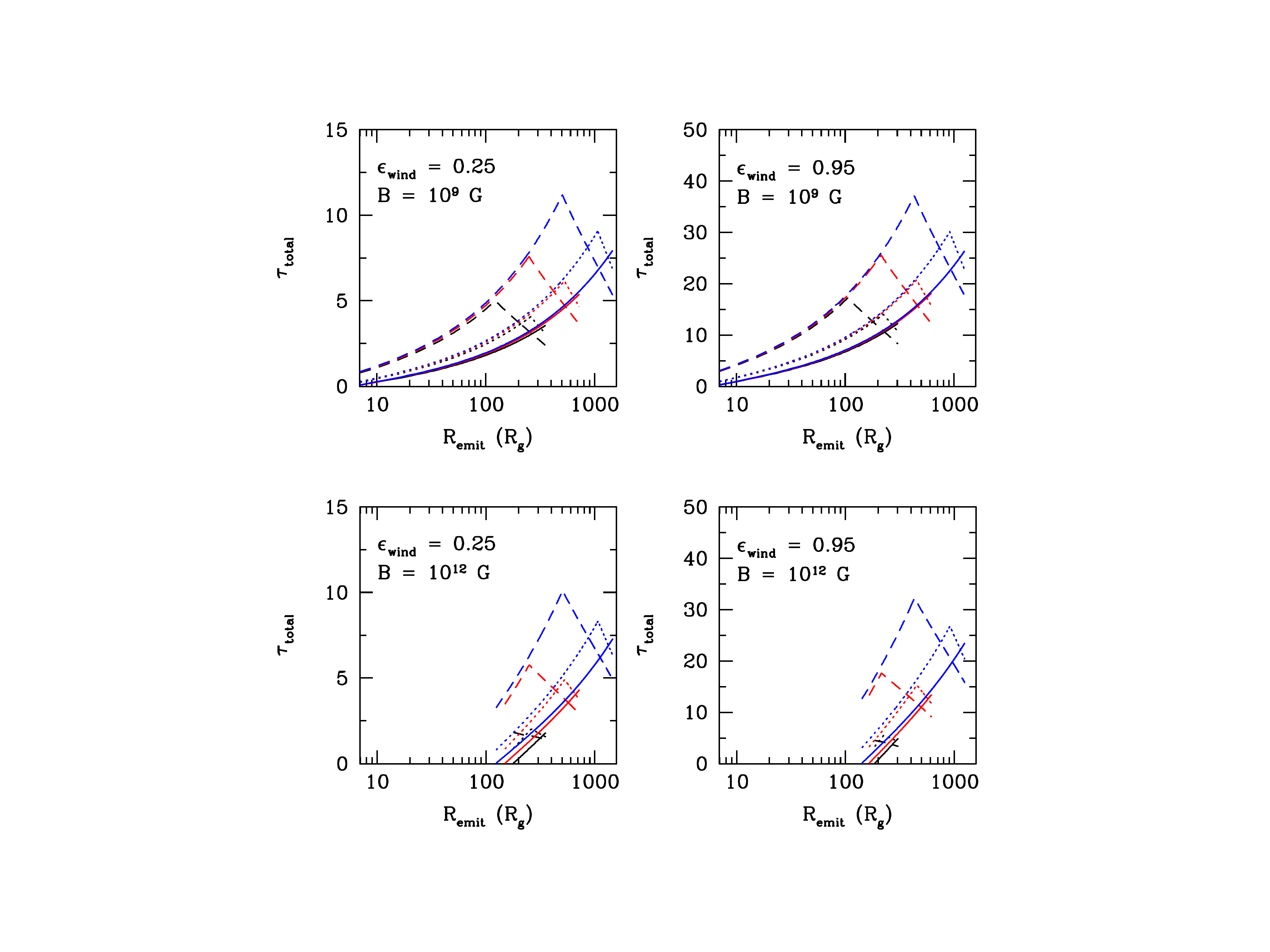}
\end{center}
\caption{As for Figure 7 but with neutron stars with dipole field strengths of 10$^{9}$ G and 10$^{12}$ G. The cut off at the magnetospheric radius  - which varies with $\dot{m}_{0}$ - is apparent.} 
\label{fig:l}
\end{figure*}


Ignoring the role of {\it outwards} advection in the wind (see Poutanen et al. 2007), the distance, $D$, through which the photons propagate (see Figure 6) can be approximated by:

\begin{equation}
D(r) \approx \frac{r}{{\rm sin}(\theta_{\rm obs})}\left[\frac{{\rm cot}(\theta_{\rm obs}) - (H/R)_{\rm disc}}{{\rm cot}(\theta_{\rm obs}) - (H/R)_{\rm wind}} - 1\right],
 \end{equation} 
 
\noindent for $\theta_{\rm obs} > 0$ (for $\theta_{\rm obs}$ = 0, $D(r)$  is simply $r[(H/R)_{\rm wind} - (H/R)_{\rm disc}]$). 

We assume a disc scale-height ($H/R_{\rm disc}$) of 0.6 (Lipunova 1999; Poutanen et al. 2007), a wind cone half-opening angle of 25\degree (such that $H/R_{\rm wind}$ = 2, to approximate the results of RMHD simulations, e.g. Sadowski et al. 2014, although see Jiang et al. 2017 for simulations at higher rates) and again assume $\beta$ = 1.4. The approximate $\tau$ for $\dot{m}_{\rm 0}$ = 50, 100 and 200 is plotted in Figures 7 \& 8 as a function of inclination ($\theta_{\rm obs}$ = 0\degree, 10\degree and 20\degree) and radius; we note that by plotting in units of $R_{\rm g}$ the resulting trends are mass-independent. In making these illustrative plots we have ignored the role of spin (we set $a_{*}$ = 0.001) and, in the case of a neutron star, we have assumed dipole field strengths of $10^{9}$ and $10^{12}$ G. We have also limited $\theta_{\rm obs} < \pi/2- \phi$ (where $\phi$ is arctan$(H/R)_{\rm wind}$) such that our line-of-sight to the innermost regions is not through the entire radial extent of the wind. Finally, we ensure that the projected radial component of the distance through the wind is not so great that when summed with $r$ it is in excess of $r_{\rm out}$ (which occurs when we are viewing through the edge of the wind). Where this occurs we set $D = (r_{\rm out}- r)/\rm{sin}(\theta_{\rm obs})$. The resulting trends for $\tau$ appear broadly self-similar (and, as expected are identical for a black hole and neutron star with B = $10^{9}$ G).

\begin{figure*}
\begin{center}
\includegraphics[trim=60 50 0 0, clip, width=12cm]{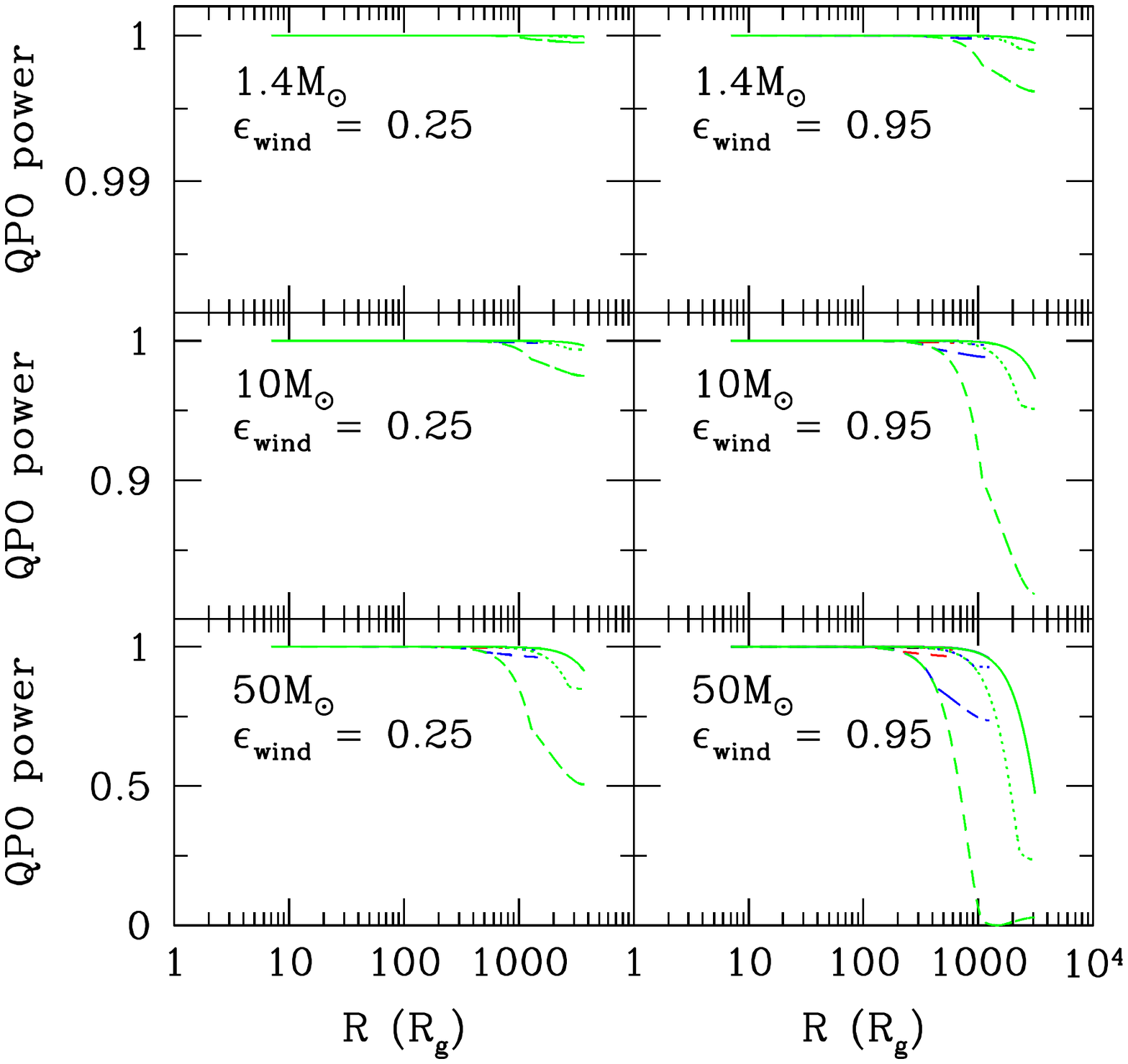}
\end{center}
\vspace{-3.5cm}
\caption{Example of the dilution of QPO variability power (with $\nu_{\rm qpo}$ = 10 mHz) as a function of compact object mass, $\epsilon_{\rm wind}$ and mean line-of-sight inclination ($\theta_{\rm obs}$) for $\dot{m}_{\rm 0}$ = 50 (black), 100 (red), 200 (blue) and 500 (green). Solid lines indicate an inclination of 0\degree, dotted  = 10\degree, dashed = 20\degree. It is clear that the QPO is unlikely to be heavily diluted unless seen at large inclinations and/or high $\dot{m}_{\rm 0}$ but even in these cases, the QPO should be readily detectable in emission from small radii - i.e. the hottest parts of the inflow, consistent with observation (e.g. Rao et al. 2010; De Marco et al. 2013).} 
\label{fig:l}
\end{figure*}


In Figure 9 we show the corresponding dilution of the QPO power by the wind as a function of $r_{\rm emit}$, for $\dot{m_{\rm 0}}$ = 50, 100, 200 and 500, across the inclination range previously used, for our bounding values of $\epsilon_{\rm wind}$ and for $\nu_{\rm qpo}$ = 10~mHz (see following section). By considering only the impact of the near-side of the outflow, we obtain an upper limit on the dilution of the QPO (as the optical depth through the outflow from the far-side will be less for $\theta_{\rm obs} \ne 0$). We have also ignored the role of dipole field strength which yields an upper limit on the dilution (as the optical depth and $D$ are largest when $r_{\rm m} \le r_{\rm isco}$ - see Figure 8). As opposed to the case of $\tau$ (Figures 7 \& 8), the dilution cannot be plotted without a mass dependence and so we show this for compact object masses of 1.4 M$_{\odot}$, 10 M$_{\odot}$ and 50 M$_{\odot}$. From this figure we can conclude that, even when the inclination, accretion rate and mass are large, a clear QPO signal is expected to originate from the innermost regions where the outflow is becoming optically thin (Poutanen et al. 2007), and from where much of the flux above 1 keV emerges (we also discuss the effect of larger inclinations than we have allowed here in the following section).

\begin{figure*}
\begin{center}
\includegraphics[trim=20 80 0 0, clip, width=18cm, angle=0]{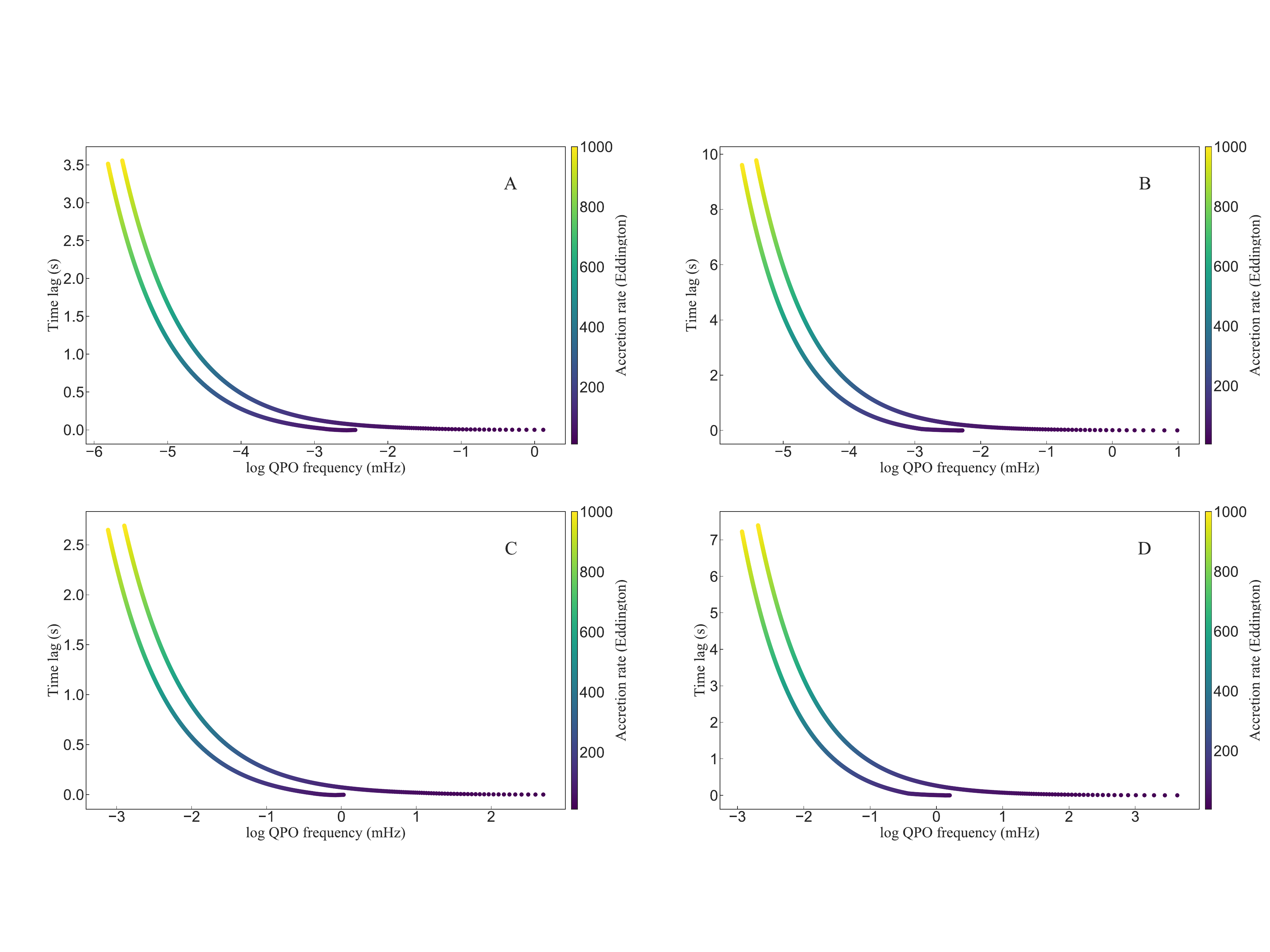}
\end{center}
\caption{Self-consistent time-lag versus QPO frequency versus accretion rate for a neutron star (M = 1.4 M$_{\odot}$) with $a_{*}$ = 0.001, $\epsilon_{\rm wind} = 0.25$ (panel A), $a_{*}$ = 0.001, $\epsilon_{\rm wind} = 0.95$ (panel B), $a_{*}$ = 0.3, $\epsilon_{\rm wind} = 0.25$ (panel C) and $a_{*}$ = 0.3, $\epsilon_{\rm wind} = 0.95$ (panel D). In each case we have chosen an inclination of 20\degree and explore the impact for example dipole field strengths of 10$^{9}$G and 10$^{12}$G (left curve and right curve in each case) numerically solving for the position of the magnetospheric radius via equations (6) and (7)} 
\label{fig:l}
\end{figure*}


\begin{figure*}
\begin{center}
\includegraphics[trim=20 80 40 0, clip, width=18cm, angle=0]{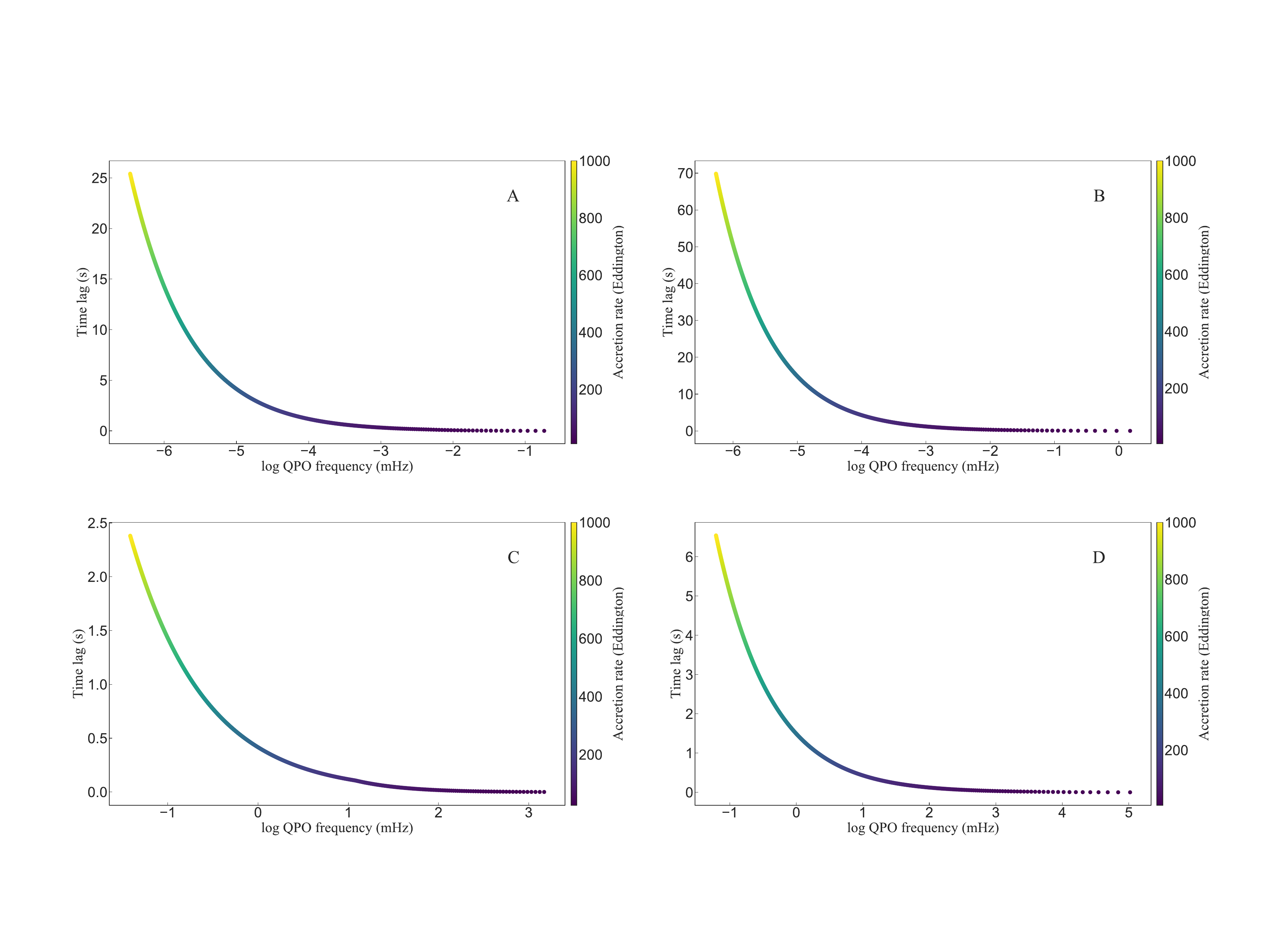}
\end{center}

\caption{Self-consistent time-lag versus QPO frequency versus accretion rate for a stellar mass black hole (M = 10 M$_{\odot}$) with $a_{*}$ = 0.001, $\epsilon_{\rm wind} = 0.25$ (panel A), $a_{*}$ = 0.001, $\epsilon_{\rm wind} = 0.95$ (panel B), $a_{*}$ = 0.998, $\epsilon_{\rm wind} = 0.25$ (panel C) and $a_{*}$ = 0.998, $\epsilon_{\rm wind} = 0.95$ (panel D). As before, we assume an inclination of 20\degree.} 
\label{fig:l}
\end{figure*}


\begin{figure*}
\begin{center}
\includegraphics[trim=20 80 0 0, clip, width=18cm, angle=0]{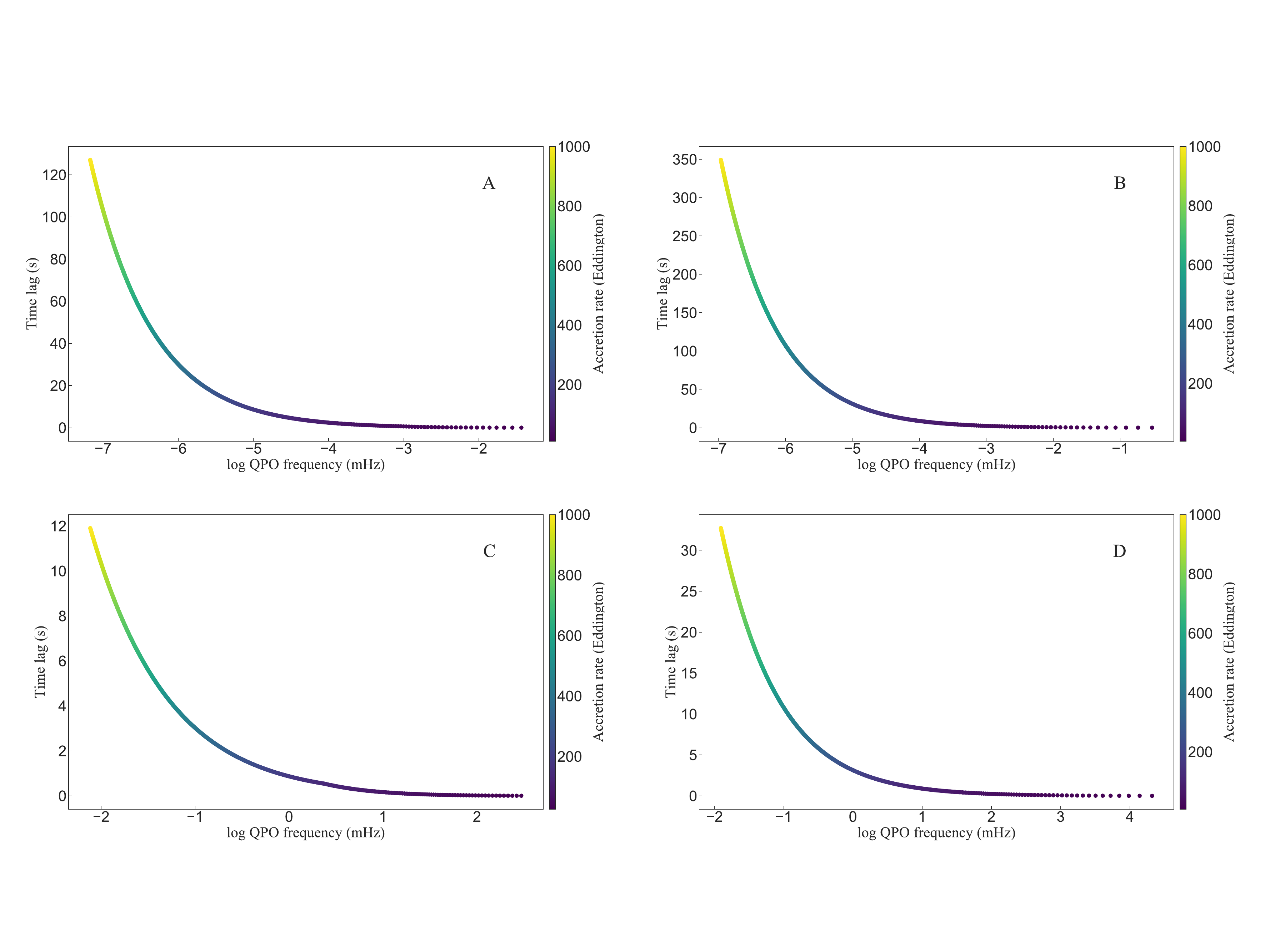}
\end{center}

\caption{Self-consistent time-lag versus QPO frequency versus accretion rate for a high mass stellar mass black hole (M = 50 M$_{\odot}$) with $a_{*}$ = 0.001, $\epsilon_{\rm wind} = 0.25$ (panel A); $a_{*}$ = 0.001, $\epsilon_{\rm wind} = 0.95$ (panel B); $a_{*}$ = 0.998, $\epsilon_{\rm wind} = 0.25$ (panel C) and $a_{*}$ = 0.998, $\epsilon_{\rm wind} = 0.95$ (panel D). As before, we assume an inclination of 20\degree.} 
\label{fig:l}
\end{figure*}


\section{Are we seeing the signature of Lense-Thirring precession in ULXs?}

The preceding sections make a series of testable hypotheses. As an example, ULXs have, to-date, detected periods/quasi-periods in the range $\sim$days to 100 days (e.g. Strohmayer 2009; Kaaret et al. 2006; Gris{\'e} et al. 2013; Motch et al. 2014; Walton et al. 2016b; Hu et al. 2017) and soft X-ray components with temperatures ranging from $\sim 0.1-0.4$ keV (Middleton et al. 2014; 2015a). Assuming that this temperature corresponds to that at the spherisation radius and using estimates for the wind parameters of $\epsilon_{\rm wind}$ = 0.25 - 0.95, $\beta$ = 1.4 and $f_{\rm col} = 2$, and equations (1), (13) and (20), we are able to infer a range in $\nu_{\rm inflow}$ for a range of mass and spin. In the case of black holes we assume spin values of 0.001 and 0.998 and for the sake of simplicity, in the case of a neutron star we have assumed that the spin is low enough such that $r_{\rm isco} = 6$. We also provide values scaled for $P_{\rm wind}$ = 1 day, whilst, in reality, $P_{\rm wind}$ is a function of spin, mass and $\dot{m}_{0}$ (via equation 3 - see Figure 4) and so the values shown in Table 1 are only to illustrate values that could be achieved after scaling by the actual observed $P_{\rm wind}$. 

\begin{table*}
\begin{center}
\begin{minipage}{150mm}
\bigskip
\caption{Predicted QPO frequencies}
\begin{tabular}{|c|c|c|c|c|c|c}

\hline 
\hline

& \multicolumn{6}{c} {$\nu_{\rm inflow}$ (mHz)}\\
$P_{\rm wind}$  & NS  & NS  & BH  & BH & BH  & BH \\
(days) & (10$^{9}$ G) & (10$^{12}$ G) & (10 M$_{\odot}$ $r_{\rm in} \approx$ 6) & (50 M$_{\odot}$ $r_{\rm in} \approx$ 6) & (10 M$_{\odot}$ $r_{\rm in} \approx$ 1.25) & (50 M$_{\odot}$ $r_{\rm in} \approx$ 1.25)\\
   \hline
   \hline
   \multicolumn{7}{c}{$\epsilon_{\rm wind}$ = 0.25}\\
   \hline
1 &  0.28 - 27.91  & 0.01 - 22.30 & 0.04 - 5.89  & X$_{\rm 2}$ - 1.57 &  0.09 - 7.14 & 0.02 - 2.06 \\   
   \hline
   \multicolumn{7}{c}{$\epsilon_{\rm wind}$ = 0.95}\\
   \hline
1  &0.56 - 100.94 & X$_{\rm 1}$ - 42.09 & 0.02 - 19.82  & X$_{\rm 2}$ - 4.67 &  0.24 - 26.70 & 0.04 - 7.47\\   
   \hline
   \hline

\end{tabular}
Notes: Approximate range in QPO frequencies derived for black holes and neutron stars (with dipole field strengths of 10$^{9}$ and 10$^{12}$~G, mass of 1.4 M$_{\odot}$, and assuming $r_{\rm isco} = 6$) for our bounding values of $\epsilon_{\rm wind}$, $T_{\rm sph}$ = 0.1 and 0.4 keV, and using equation (2). The input wind precession period is 1 day such that the rough predicted QPO period can be found by factoring through by the observed value in the same units. X$_{\rm 1}$ refers to the case where $r_{\rm m} \ge r_{\rm sph}$ and X$_{\rm 2}$ the case where the accretion rate is so low that the photosphere sits within the spherisation radius.
\end{minipage} 
\end{center}
\end{table*}

QPOs at $\sim$ 10s of mHz are seen in the ULXs NGC 5408 X-1 (Strohmayer \& Mushotzky 2009; Pasham et al. 2012), NGC 6946 X-1 (Rao et al. 2010), NGC 1313 X-1 (Pasham et al. 2015), M82 X-1 (Strohmayer \& Mushotzky 2003 - although this may yet be a good candidate for a fairly high mass stellar mass black hole - Brightman et al. 2016 - or even IMBH - Pasham, Strohmayer \& Mushotzky 2014) and claimed in IC 342 X-1 (Agrawal \& Nandi 2015). Although longer timescale periodic/quasi-periodic variations (simultaneous with the QPO detection) have yet to be identified in all of these sources, it is apparent that certain combinations of physical parameters would allow their QPOs to be explained by our Lense-Thirring model (noting once again that there are a number of uncertainties, e.g. the location of the relevant radii - see \S 5 for more details). We note that, whilst QPOs at similar frequencies have been seen in other accreting systems (e.g. LMXBs, e.g. Altamirano \& Strohmayer 2012), these are not accreting at super-critical rates and we are not implying that our model is an appropriate mechanism in such cases.

It is worth noting that, where energy-dependent analyses have been performed, the mHz QPOs in ULXs appear to have the highest variability power in the hard X-rays (e.g. Rao et al. 2010; De Marco et al. 2013) consistent with our predictions regarding the energy dependence of signal dilution due to propagation (Figure 9). Given the predicted geometrical nature of the QPO, we should also see only a weaker signature of the precessing inflow when the source is face-on (with a small precessional cone) but with increasing inclination (or large precessional cone), the QPO should appear stronger (an analogous situation has been inferred for low frequency QPOs in X-ray binaries - Motta et al. 2015). At sufficiently high inclinations, the line-of-sight optical depth will likely include the much denser inflow and, for large masses and accretion rates, may be too large to allow the QPO to be detected (see Figure 9). Based on the spectral-timing properties (Middleton et al. 2015a) and the detection of mass-loaded winds via atomic absorption lines (Middleton et al. 2014, 2015b; Pinto et al. 2016; Walton et al. 2016a), it is perhaps unsurprising that three of the ULXs where QPOs have been firmly detected (NGC 5408 X-1, NGC 6946 X-1 and NGC 1313 X-1) are consistent with being viewed at moderate inclinations to the line of sight - ideal conditions for detecting a mHz QPO due to precession.

\subsection{QPO time lags}

As a consequence of propagation through the wind, it is inevitable that there should be a time-lag imprinted at the QPO frequency between well-separated energy bands corresponding to the innermost, least optically thick, hot regions and bands where emission from the spherisation radius emerges (noting that the former will not extend down to $r_{\rm isco}$ where the compact object is a neutron star with a dipole magnetic field strength $> 10^{9}$ G). Given the lower optical depth to the inner-most regions at most inclinations (that do not extend to sight-lines through the entire extent of the wind), it is likely that this will take the form of a `soft-lag' (i.e. the soft emission lagging the hard emission). Although we do not explore this region of parameter space here, we note that at inclinations into the wind itself (i.e. $\theta_{\rm obs} \ge \theta_{\rm wind}$), the distance to the innermost regions extends through the entire extent of the wind leading to extensive Compton-down scattering (such that the QPO from these regions will be heavily diminished and any residual signal will emerge at a very different energy) and the QPO signal from the outermost regions may start to lead that from smaller radii (i.e. a hard lag will be produced).

As can be seen from equation (2), the QPO frequency is a function of the various system parameters (mass, spin and accretion rate) from which the lag is also inferred and which provides a further testable, unique hypothesis for this model. As an example, we plot the values for the {\it maximum} lag (i.e. that between the hard emission from $r_{\rm in}$ and the soft emission from $r_{\rm sph}$ considering only the near-side of the outflow) at the corresponding QPO frequency for a neutron star with a mass of 1.4 M$_{\odot}$ (and example dipole field strengths of 10$^{9}$ and 10$^{12}$ G - Figure 10) and for black holes of 10 M$_{\odot}$ and 50 M$_{\odot}$ (Figures 11 \& 12). We assume a range in $\dot{m}_{\rm 0}$ of 5-1000, an example inclination of 20\degree and $\epsilon_{\rm wind}$ = 0.25 and 0.95. In the case of a neutron star, we use spin values of 0.001 and 0.3, and for the black holes, spin values of 0.001 and 0.998.

\subsection{NGC 5408 X-1}

To date, the strongest claim of a soft-lag from a ULX is that for NGC 5408 X-1 (see Heil \& Vaughan 2010, De Marco et al. 2013, Hernandez-Garcia et al. 2015) with QPOs observed between 10 and 38 mHz (De Marco et al. 2013) with the soft band (0.3-1 keV) lagging the hard band (1-7 keV) by $\approx$ 1.5-5s at the QPO frequency (we note that soft lags have been reported in other ULXs at candidate QPO frequencies - Li et al. 2017). Spectral fitting using a multi-colour disc blackbody component (in {\sc xspec} this is through use of the {\sc diskbb} set of models which do not include a colour temperature correction: Mitsuda et al. 1984) implies the temperature of the soft X-ray component in NGC 5408 X-1 lies between $\approx$ 0.12 - 0.23 keV (see Kaaret et al. 2003; Soria et al. 2004; Strohmayer \& Mushotzky 2007; Middleton et al. 2014; 2015a). For the observed range in $T_{\rm sph}$, assuming $f_{\rm col} = 2$ and a range in $\dot{m}_{\rm 0}$ of 5-1000, we can infer approximate bounds on the mass and spin of the compact object, accretion rate through the outer disc and mean inclination to the line-of-sight assuming the QPO originates from Lense-Thirring precession (equation 2) and accepting the caveats associated with our model. 

We note that in comparing our analytical model with observation, we are making an explicit assumption that the value reported from the Fourier lag spectrum is equivalent to the lag in our model. This may be inaccurate for a number of reasons. Firstly, the true lag measured from the cross spectrum (see the review of Uttley et al. 2014) is obtained from the phase of the product of the driving signal (in this case the QPO and underlying broad-band noise) and the transfer function. In addition, an accurate picture requires the radial contribution to the lag (i.e. the energy dependence) and impact of changing inclination with QPO phase to be determined, as well as the impulse response function to be evaluated over {\it both} azimuth and inclination. These complications are beyond the scope of this initial work, however, we note that accounting for the contribution from all azimuths is likely to reduce the lag in the model for a given inclination (other than face-on), mass and accretion rate, as the distance to the inflow through the wind ($D$ in Figure 6) is smaller (and the optical depth is lower via equation 22). In addition, a range of radii must contribute to each energy band, which will dilute the predicted lag still further. At lower frequencies, we expect a lag to hard energies in the broad-band noise due to propagation of fluctuations through the inflow (e.g. Lyubarskii 1997; Ar{\`e}valo \& Uttley 2006; Ingram \& Done 2012; Hernandez-Garcia et al. 2015), possibly affected by mass loss in the wind (see Middleton et al. 2015a). As the lag due to scattering through the wind is super-posed onto this, it is possible that the lag we are using to determine the properties of the compact object is being somewhat diluted which would then affect resulting parameter estimates. Although not included in our simple model, in future we will extract the relevant (hard and soft) lags directly from GRMHD simulations, both at the QPO frequency and across the broad-band noise.

Noting the above caveats, we proceed to apply our model assuming $\zeta$ = 2 and $\epsilon_{\rm wind}$ = 0.95; in using the latter value as a reasonable upper limit to the true value of $\epsilon_{\rm wind}$, we further constrain the approximate {\it lower} limit on the mass. Initially we test the case where the compact object is a black hole with a mass in the range 3-50 M$_{\odot}$ finding that masses $\ge$ 11 M$_{\odot}$, high spins ($a_{*}>$0.8) are preferred, with the system viewed at grazing inclinations to the wind, and $\dot{m}_{\rm 0}$ between 40 and 162 (see Figure 13).

Noting that the accretion rate and mass are covariant (e.g. equations 2 \& 19), we also determine the range in intrinsic and maximum (beamed) luminosities from $L_{\rm int} \approx L_{\rm Edd}\left[1+{\rm ln}(\dot{m}_{0})\right]$ (which is appropriate for a black hole system, in the case of a neutron star, a slightly different formula should be used: Erkut et al. 2019) and $L_{\max} \approx L_{\rm int}\dot{m}_{0}^{2}/73$ (King 2009). The luminosity range we would infer, ranges from  $\approx 9\times10^{39}$ erg/s - 4$\times10^{40}$ erg/s (intrinsic) with corresponding beamed luminosities of 3$\times10^{42}$ erg/s - 4$\times10^{42}$ erg/s. However, it is important to note that some component of the intrinsic luminosity must be used to launch the wind (Poutanen et al. 2007), and only the remaining radiative power can then be beamed (thus our values are firm upper limits). Given that a fraction (which is potentially large: Pinto et al. 2016) of the available power is lost to the wind, then the observed (X-ray) luminosity of the system at $\gtrsim$ 5$\times10^{39}$ erg/s (Middleton et al. 2015a) is consistent with the values we have inferred. 

As we do not yet know the system parameters for NGC 5408 X-1 (companion mass or orbital period), we cannot yet say for certain whether the Lense-Thirring torque will dominate over the tidal torque; however, by inspection of Figure 2, it would appear that there is ample range to allow the precession of the wind to occur unhindered by the presence of the companion star (likely to be a super-giant: Gris{\'e} et al. 2012).

We test for the presence of a neutron star separately for the same range in the accretion rate and disc/wind parameters, at dipole field strengths of 10$^{9}$ - 10$^{12}$ G (in steps of $\times$10 G), but find that, in all cases where the QPO frequency can be matched, the lag is too small. Our simple model therefore appears to exclude a neutron star primary from NGC 5408 X-1. Although our model is still in its infancy, as discussed above, we do not expect lower masses to be compatible after considering contributions from all azimuths, however, as we will discuss, including additional opacities may yet allow lower masses to be reached. 

\begin{figure*}
\begin{center}
\includegraphics[trim=0 100 0 150, clip, width=16cm]{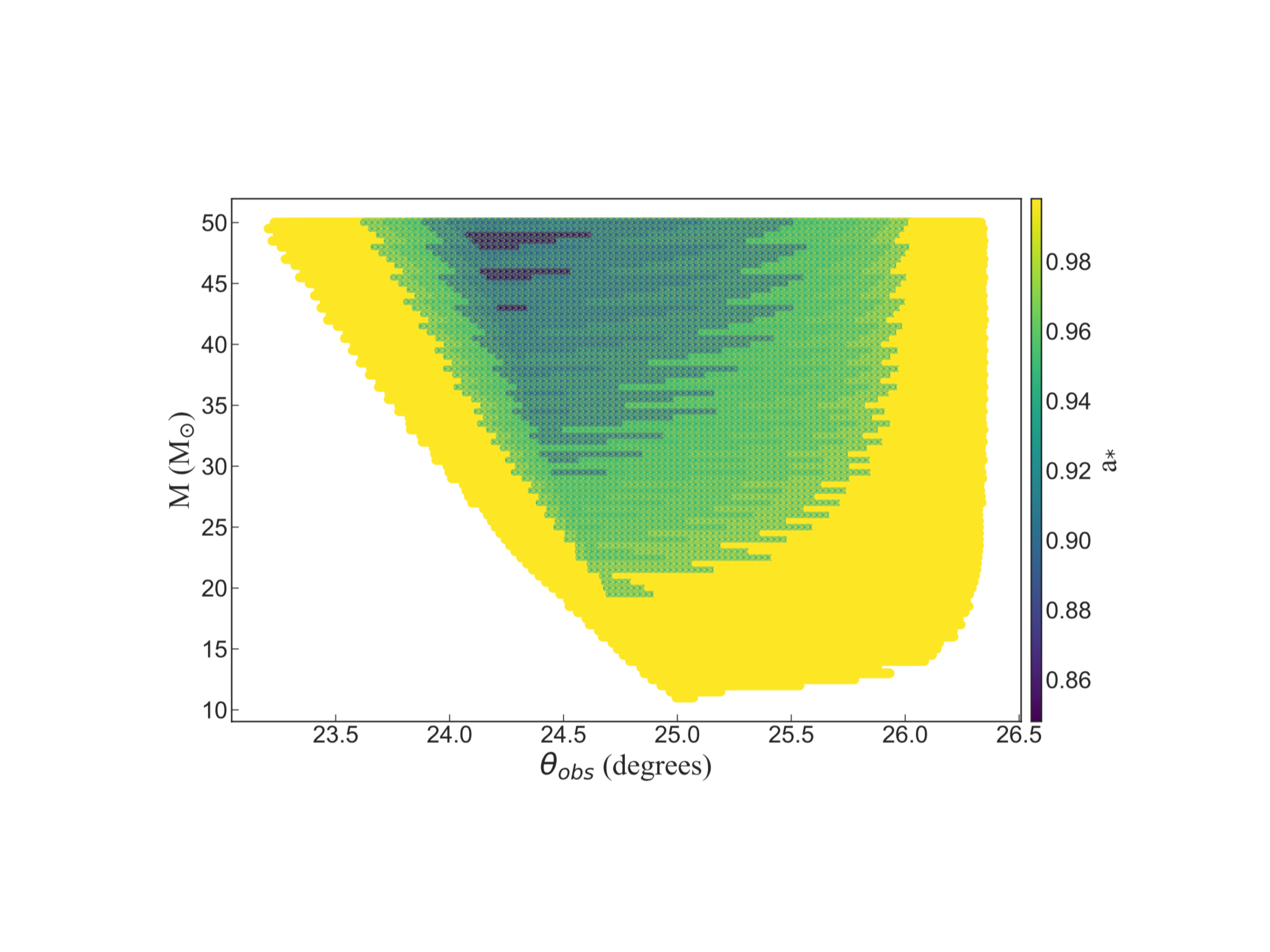}
\end{center}
\vspace{-0.2cm}
\caption{Mass, spin and inclination parameter space for NGC 5408 X-1 for a QPO lying between 10 and 38~mHz (De Marco et al. 2013), $T_{\rm sph}$ between 0.12 and 0.23~keV, and for lags between 1.5 and 5~s in duration (De Marco et al. 2013). Based on our simple, analytical model (assuming H/R$_{\rm wind}$ = 2), we would expect NGC 5408 X-1 to contain a highly spinning, fairly massive stellar mass black hole ($M \ge 11 M_{\odot}$) seen at grazing angles to the wind (however, as we note in the main text, this may produce a tension with the X-ray spectrum, the solution to which likely requires the diagnostic power of a full GRMHD treatment).} 
\label{fig:l}
\end{figure*}

Using equation (3), we also determine the range of wind precession periods for NGC 5408 X-1 that could accompany the compatible range of physical values from applying our model above. These extend from $\approx$ 4 hours to $\approx$ 1 day (the latter occurring for the lowest mass, highest spin and highest accretion rate). Whilst 100+ day periods have been claimed in this source (e.g. Strohmayer 2009), we note that the strongest candidate period is actually 2.7 days (Gris{\'e} et al. 2013; An et al. 2016). This period has been shown to be transient and therefore cannot be the orbital period of the system, however, if this is instead associated with the wind precession then variations in the mass accretion rate readily lead to changes in the precession period (equations 2 and 3) and can even damp the precession entirely should the spherisation radius grow too large (Motta et al. 2018). Although 2.7 days is slightly longer than our simple model would predict, this is not unexpected in light of the many caveats discussed so far - and it is certainly conceivable that errors in the uncertainty in the location of the various radii, the impact of radiative torques within the flow itself or the lack of strict simultaneity between the QPOs/lags and the strongest period determination (cf Gris{\'e} et al. 2013; De Marco et al. 2013) might account for such a discrepancy.

Estimates for the physical parameters via our current simple model may be influenced by the role of bound-free absorption which we have hitherto ignored. This can be somewhat justified as the densities and temperatures we have inferred from the analytical model (under the assumption that the temperature of the wind material is that of the inflow at that radius), imply that Thompson opacity should be dominant. However, the outflowing material will no doubt cool and, given the steep dependence of Kramers opacity on the temperature, it is plausible that the associated optical depth could increase. Given a more physically accurate description of the wind, where self-shielding may be important, it is therefore possible that the lag could also increase. Certainly, observations imply that the column density associated with neutral gas in the wind may be potentially large (see Middleton et al. 2015b), with additional resonance line opacities also contributing (Pinto et al. 2016; 2017; Walton et al. 2016). Together with a lag due to recombination (see Silva et al. 2016), our model may somewhat overestimate the mass of the compact object; such effects will be studied in detail in future work.

In addition to coherent power in the form of QPOs, we note that there may also be a lag for all variability generated in the inner regions propagating through the wind. The short timescale ($<$ 1000s seconds) broad-band variability in NGC 5408 X-1 (which dominates the power in these softer ULXs - Sutton, Roberts \& Middleton 2013; Middleton et al. 2015a) has been proposed to arise as a result of obscuration by optically thick clumps (Middleton et al. 2011, 2015a), assumed to be generated by radiative-hydrodynamic instabilities (e.g. Takeuchi et al. 2013) in the wind which we know to be present in this source (Pinto et al. 2016, see also Middleton et al. 2015b). As this variability is {\it extrinsic} in nature - imprinted by the wind itself - a lag of similar origin to that of the QPO (i.e. propagation through intervening material) could in principle be created, however, this will depend on where the variability is imprinted, e.g. by denser material at larger radii. A detailed model is beyond the scope of this current work but will be presented in future.

\section{Discussion \& Conclusions}

By associating the $\sim$day-timescale periods and mHz QPOs seen in ULXs, with Lense-Thirring precession of super-critical accretion flows, we are able to make a series of distinct predictions. Foremost is that, when plotted against the temperature of the soft X-ray component ($T_{\rm sph}$), we are able to map out planes with distinct regions in which black hole ULXs might be preferentially or exclusively found. In addition, the ratio of precession frequencies also forms an accretion plane which is mostly insensitive to the magnetic dipole field strength of NS ULXs at high $\dot{m}_{0}$. If $\epsilon_{\rm wind}$ is found to be low then the regions corresponding to black holes and neutron stars are well separated, however, if it is instead high, then indirect constraints on the dipole magnetic field may be required to use this plane fully. As pulsations are inherently transient (Bachetti et al. 2014) and high rates of accretion may well have suppressed the surface field in many neutron star ULXs (such that pulsations and electron/proton resonance scattering features will be absent), these planes of accretion may provide one of the few means to separate the population of black holes and neutron stars in ULXs (see the related discussions of King \& Lasota 2016; Middleton \& King 2017; Walton et al. 2018b). We note that, as the Lense-Thirring precession periods are functions of accretion rate through the outer disc, any `noise' (variations in $\dot{m}_{\rm 0}$) will lead to variations in the periods - they become quasi-periodic oscillations. It is therefore important that, when measuring the ratio of day timescale period to mHz QPO that they are contemporaneous. Where the variation in $\dot{m}_{\rm 0}$ is minimal or can be accounted for, secular evolution in the periods will also indicate the presence of a neutron star, as the spin (and the associated precession periods) will not evolve on observational timescales where the compact object is a black hole. This will be true irrespective of whether the ULX is Lense-Thirring precessing or if the signal is instead due to precession of the magnetic dipole field (e.g. Lipunov \& Shakura 1980; Mushtukov et al. 2017). 

Day timescale and mHz QPOs are already detectable in the bright ULXs and the prospects for further discovery will improve substantially in the era of high throughput missions such as {\it Athena} (and {\it LOFT} type missions such as {\it STROBE-X}). Thus, in future, we can hope to better test our predictions against a larger sample of objects and use these timing signals to infer the mass/spin range of compact objects in ULXs; this is in stark contrast to determining the mass via dynamical means which has been shown to be challenging at best (e.g. Roberts et al. 2011 but also Liu et al. 2013; Motch et al. 2014).

In addition to the prediction of mass segregation from use of the precession periods, our model provides a self-consistent explanation for the time-lag at the mHz QPO frequency seen in the archetypal ULX, NGC 5408 X-1 (see De Marco et al. 2013, Hernandez-Garcia et al. 2015) as well as predictions for the relationship between QPO frequency and lag. The production of a lag in this model is due entirely to the effects of propagation as opposed to thermal reverberation (De Marco et al. 2013) and we have ignored the effects of light travel time/reprocessing. This is reasonable as the maximum reverberation lag possible in our model is given by the additional path length introduced by scattering from the far side of the optically thick wind, i.e. $t_{\rm lag} \le \sqrt{2}R_{\rm out}/c$. Applying our simple model to NGC 5408 X-1 implies the compact object is a stellar mass black hole with a high spin, with reverberation lags (determined from the various combinations of black hole mass and accretion rate) of $<$ 1 second, smaller than the lags we are considering here. Certainly the environment of NGC 5408 X-1 has a low metallicity (Mendes de Oliveira et al. 2006, Gris{\'e} et al. 2012) which would tend to favour the formation of black holes due to low mass-loss from the progenitor. Combined with the possibility of high spin, the Blandford-Znajek mechanism (Blandford \& Znajek 1977) may then contribute to the production of powerful jets. Indeed, whilst ejections have yet to be confirmed in this source, a bright radio counterpart has been detected by Kaaret et al. (2003) with a flux of $\approx$~0.26~mJy at 4.8~GHz, which would be enormously bright if located in our Galaxy ($\approx$ 50 Jy at a distance of 10 kpc). Although jet ejections have not yet been resolved for NGC 5408 X-1, we note that extremely powerful ejections {\it have} been resolved from Ho II X-1 (Cseh et al. 2014; 2015) which shares many similarities with NGC 5408 X-1 (e.g. both show soft X-ray spectra: Middleton et al. 2015a, absorption features: Middleton et al. 2015b and both are surrounded by radio nebulae: Cseh et al. 2012). Although we are able to explain key elements of the phenomenology of NGC 5408 X-1 with our model, we note that the X-ray spectrum has been ascribed to a viewing angle somewhat into the wind in order to suppress the luminosity of the hard X-ray component (see Middleton et al. 2015a) and explain the presence of absorption features (Middleton et al. 2014; 2015b, Pinto et al. 2016). Whilst this would not be immediately consistent with the grazing angles we infer, it is important to stress that we have not yet incorporated a physical model for the wind which may allow this tension to be resolved.

There are naturally a number of additional caveats to this work, many of which will be explored in future via GRMHD simulations. Regarding our assumed dominance of Lense-Thirring torques, we have not explicitly considered the role of magnetic torques - this is reasonable if the ULX contains a black hole or if the neutron star is not highly magnetised, as may well be the case after prolonged accretion at such high rates (Bhattacharya 2002) and/or if the disc is thick (Middleton et al. 2018). We have also not considered the role of torques resulting from the (non-axisymmetric) radiation field from the precessing flow beneath the wind; this is a complex issue, the detailed understanding of which cannot be explored without simulations which do not presently exist. Regarding values entering into our calculations, we have assumed the location of the relevant radii are consistent with the analytical formulae (Poutanen et al. 2007) and that the wind precesses out to $r_{out}$. We have also assumed that the properties of the wind are independent of spin (and, in the case of neutron stars, independent of the dipole field strength), and that the wind velocity is simply equal to the local escape velocity. Finally, we have assumed that the temperature we infer from the soft X-ray spectral component can be associated directly with the spherisation radius whilst this will no doubt be affected by any scattering and thermalisation of incident harder X-rays from smaller radii (and irradiation by an accretion column or the neutron star surface - see Mushtukov et al. 2017). Regarding the time-lag calculations, at present, the model  does not account for the contribution to the lag from all azimuths and ignores the role of bound-free opacity. Naturally as we develop our model based on insights derived from the latest simulations (e.g. Sadowski et al. 2014; Jiang et al. 2014, 2017) which do not rely on the average properties of the wind nor treat it as a laminar outflow, we will be able to more rigourously test the theoretical predictions we have made here.

\section{Acknowledgements}

The authors thank the anonymous referee for their helpful suggestions. MJM appreciates support from an Ernest Rutherford STFC fellowship. PCF appreciates support from National Science Foundation grants AST1616185 and PHY-1748958. AI appreciates support from the Royal Society. TPR appreciates support via STFC consolidated grant ST/L00075X/1. We thank Omer Blaes, Alexander Mushtukov and Rajath Sathyaprakash for helpful discussion and suggestions. 

\label{lastpage}

\end{document}